\documentclass[jou]{interact}

\usepackage{dsfont}
\usepackage{xcolor}
\usepackage{tcolorbox}
\usepackage{cancel}

\usepackage{amsfonts}
\usepackage{fancyvrb}
\usepackage[english]{babel}
\usepackage{inputenc}

\usepackage[notocbib]{apacite} 

\usepackage{multicol, amsthm}

\usepackage{bbm}
\usepackage{courier}

\usepackage{parskip}
\usepackage{scalerel,stackengine}
\stackMath
\newcommand\reallywidehat[1]{%
\savestack{\tmpbox}{\stretchto{%
  \scaleto{%
    \scalerel*[\widthof{\ensuremath{#1}}]{\kern-.6pt\bigwedge\kern-.6pt}%
    {\rule[-\textheight/2]{1ex}{\textheight}}%WIDTH-LIMITED BIG WEDGE
  }{\textheight}% 
}{0.5ex}}%
\stackon[1pt]{#1}{\tmpbox}%
}

\usepackage{booktabs,array}

\newcount\rowc

\makeatletter
\def\ttabular{%
\hbox\bgroup
\let\\\cr
\def\rulea{\ifnum\rowc=\@ne \hrule height 1.3pt \fi}
\def\ruleb{
\ifnum\rowc=1\hrule height 1.3pt \else
\ifnum\rowc=6\hrule height \heavyrulewidth 
   \else \hrule height \lightrulewidth\fi\fi}
\valign\bgroup
\global\rowc\@ne
\rulea
\hbox to 10em{\strut \hfill##\hfill}%
\ruleb
&&%
\global\advance\rowc\@ne
\hbox to 10em{\strut\hfill##\hfill}%
\ruleb
\cr}
\def\endttabular{%
\crcr\egroup\egroup}

\usepackage{epstopdf}% To incorporate .eps illustrations using PDFLaTeX, etc.
\usepackage[caption=false]{subfig}% Support for small, `sub' figures and tables

\usepackage[doublespacing]{setspace}% To produce a `double spaced' document if required
\usepackage{inputenc}

\theoremstyle{plain}% Theorem-like structures provided by amsthm.sty

\theoremstyle{definition}

\theoremstyle{remark}

\usepackage{natbib}
\begin{document}
\bibliographystyle{apacite}

\articletype{Original Research Paper}% Specify the article type or omit as appropriate

\title{ {Equivalence testing for linear regression}}

\author{
\name{Harlan Campbell}
%\name{XXXX}
\affil{University of British Columbia, Department of Statistics \\
Vancouver, British Columbia, Canada\\
harlan.campbell@stat.ubc.ca}
}

\maketitle

\begin{abstract}
 {We introduce equivalence testing procedures for linear regression analyses. Such tests can be very useful for confirming the lack of a meaningful association between a continuous outcome and a continuous or binary predictor.  Specifically, we propose an equivalence test for unstandardized regression coefficients and an equivalence test for semipartial correlation coefficients.  We review how to define valid hypotheses, how to calculate $p$-values, and how these tests compare to an alternative Bayesian approach with applications to examples in the literature.}
\end{abstract}

\begin{footnotesize}

\paragraph*{Author notes - }

%This manuscript has been posted as a pre-print on: \url{https://arxiv.org/abs/2004.01757} ; see \citet{campbell2020equivalence}.  
Thank you to Prof. Paul Gustafson for the helpful advice with preliminary drafts and thank you also to Prof. Dani\"{e}l Lakens and Prof. Ken Kelley for feedback and insights.

\end{footnotesize}

\newpage

\section{Introduction}

All too often researchers will conclude that the effect of an explanatory variable, $X$, on an outcome variable, $Y$, is absent when a null-hypothesis significance test (NHST) yields a non-significant $p$-value (e.g., when the $p$-value $> 0.05$).  Unfortunately, such a procedure is logically flawed.  As the saying goes, ``absence of evidence is not evidence of absence'' \citep{hartung1983absence, altman1995statistics}.  Indeed, a non-significant result can instead be due to insufficient statistical power, and while a NHST can provide evidence to \emph{reject} the null hypothesis, it cannot provide evidence to \emph{accept} the null.  

To properly conclude that an association between $X$ and $Y$ is absent or at most negligible (i.e., to confirm the \emph{lack} of an association), the recommended frequentist tool, the equivalence test (also known as the ``non-inferiority test'' for one-sided testing), is well-suited \citep{wellek2010testing}.  Let $\theta$ be the parameter of interest representing the association between $X$ and $Y$.  An equivalence test reverses the question that is asked in a NHST.  Instead of asking whether we can reject the null hypothesis of no effect, i.e., reject $\textrm{H}_{0}: \theta = 0$, an equivalence test examines whether the magnitude of $\theta$ is at all meaningful by asking: {Can we reject the possibility that $\theta$ is as large or larger than our smallest effect size of interest, $\Delta$?}   The null hypothesis for an equivalence test can therefore be defined as $\textrm{H}_{0}: \theta \notin (-\Delta, \Delta)$.  In other words, \emph{equivalence} implies that $\theta$ is small enough that any non-zero effect would be at most equal to $\Delta$.  To be clear, the interval $(-\Delta, \Delta)$ is known as the ``equivalence margin'' and represents the range of values for which $\theta$ can be considered negligible.

%and \citet{niazi2007handbook} provides a summary of the continuously growing literature on the subject  ;;;; \cite{ball2013beyond} who consider how equivalence tests may be useful for effectively testing the so-called ``gender similarities hypothesis.'' 

%\cite{marcoulides2017new} consider equivalence testing to assess structural equation models (i.e., for model fit).

{Statistical methods for equivalence testing have their origins in the 1970s and 1980s (e.g., \citet{westlake1972use},   \citet{schuirmann1987comparison}, \citet{anderson1983new}).   In psychology research and in the social sciences more broadly, the practice of equivalence testing is relatively new but is ``rapidly expanding'' \citep{koh2013robust}.  Recent examples of equivalence testing in the applied psychological research literature include \citet{fruehauf2021cognitive} who use equivalence testing to study cognitive control in obsessive-compulsive disorder, and \citet{leonidaki2021comparison} who use equivalence testing in a study of cognitive behavioural therapy.}

{Statistical software for equivalence testing is also rapidly expanding.  For researchers using R, the packages ``equivalence'' and ``TOST''  \citep{robinson2016package, lakens2017equivalence} provide many accessible functions.  For researchers using SAS, STATA, or SPSS, there are also many available resources; \citet{batterham2016improved} (in their Appendix) provide a summary with examples.}

{
In the psychological research methods literature, one early appearance of equivalence testing methods is \citet{rogers1993using} who discuss equivalence testing for group mean differences.  {More recent examples include \citet{goertzen2010detecting} who highlight the importance of using equivalence tests to establish the independence of two different variables (i.e., for establishing negligible correlations), \citet{counsell2020evaluating} who consider equivalence testing methods for measurement invariance, and \cite{marcoulides2017new} who consider equivalence testing for assessing structural equation models.  Outside of psychology there is also a growing literature on equivalence testing methods (e.g.,  \citet{leday2022multivariate} and \citet{shen2023equivalence}).}}

{Conspicuously absent is any published research on methods for equivalence testing in linear regression, with the notable exception of \citet{dixon2005statistical} who propose equivalence tests for establishing negligible population trends in ecology.  This is rather surprising since linear regression is arguably one of the most commonly used methods for statistical analysis.  The first objective of this paper is therefore to address this research gap by outlining a general equivalence testing method for establishing negligible regression coefficients.  In Section 2 (``Equivalence testing for unstandardized regression coefficients''), we review how to define valid hypotheses, calculate $p$-values, and establish ``equivalence confidence intervals'' \citep{seaman1998equivalence} for equivalence tests of unstandardized regression coefficients.} 

{Despite becoming more common and despite the fact that available software has made it more accessible, equivalence testing remains challenging for many researchers.}  Specifically, defining and justifying the equivalence margin is cited as one of the ``most difficult issues'' \citep{hung2005regulatory}.   \citet{lakens2018equivalence} provide some guidance for using equivalence tests in psychological research but note that defining the margin will be the ``biggest challenge for researchers'' because psychological theories are often ``too vague.''  If the equivalence margin is too large, any claim of equivalence will be considered meaningless. On the other hand, if the margin is somehow too small, the probability of declaring equivalence will be substantially reduced \citep{wiens2002choosing,keefe2013defining, campbell2018make}. 
%While the margin is ideally based on some objective criteria, these can be difficult to justify, and there is generally no clear consensus among stakeholders \citep{keefe2013defining, campbell2018make}.  

Scores from many psychological measures/scales are interpretable and meaningful, and researchers should, whenever possible,  use validated and well-scaled measures where the units of measurement are well understood.  However, in certain scenarios, the parameters of interest are measured on different and somewhat arbitrary scales.  This makes the task of defining the equivalence margin more challenging.  Without units of measurement that are easy to interpret, defining and justifying an appropriate equivalence margin can be all but impossible \citep{lakens2018equivalence}. 

%(???) Moreover, without the ability of researchers to conduct equivalence testing with standardized effect sizes,  may facilitate the adoption of ``default'' or generally accepted norms for acceptable equivalence margins.(???) ''clear consensus''

{When working with parameters measured on arbitrary scales (e.g., Likert scales), researchers will often prefer to work with standardized effect sizes to aid with interpretation \citep{wilkinson1999statistical, baguley2009standardized}.  It therefore stands to reason that, for equivalence testing in such a situation, it would also be preferable to define the equivalence margin in terms of a standardized effect size.  For linear regression analyses, reporting standardized effect sizes is quite common \citep{bring1994standardize, west2007multiple} and the semipartial correlation coefficient is a standardized effect size that can be easily interpreted \citep{dudgeon2016comparative}.  Therefore, our objective in Section 3 (``Equivalence testing for a standardized effect size in linear regression'') is to establish an equivalence test for the semipartial correlation coefficient.}

%In order to evaluate the operating characteristics of our proposed test, we conduct a small simulation study.
%{We note that the tests of standardized effect sizes we propose are derived from inverting the noncentrality parameter confidence intervals put forward by \cite{kelley2007confidence} who derived these confidence intervals by pivoting cumulative distribution functions.}}

{Several Bayesian methods (e.g.,  \citet{morey2011bayes}, \citet{rouder2012default}, \citet{bedrick2018approach}) have been proposed for establishing equivalence.} {While the focus of this paper is frequentist equivalence testing, in Section 4 (``A Bayesian alternative for establishing equivalence in a linear regression''), we briefly review one of the proposed Bayesian alternatives.}

{Finally, in Section 5 (``Practical Examples''), we demonstrate how all of the different testing methods can be applied in practice with {two} practical examples.  We then conclude in Section 6 with some general recommendations on how to perform equivalence testing for linear regression.  In the supplemental material, R code is available to implement all of the calculations and analyses.}
 
\section{{Equivalence testing for unstandardized regression coefficients}}
\label{sec:R2}
%
%An equivalence test reverses the question that is asked in a null hypothesis significance test (NHST).  Instead of asking whether we can reject the null hypothesis of no effect, e.g., $\textrm{H}_{0}: \theta = 0$, an equivalence test examines whether the magnitude of $\theta$ is at all meaningful: \emph{Can we reject the possibility that $\theta$ is as large or larger than our smallest effect size of interest, $\Delta$?}   
%
%The null hypothesis for an equivalence test is defined as $\textrm{H}_{0}: \theta \notin [-\Delta, \Delta)$.  In other words, \emph{equivalence} implies that $\theta$ is small enough that any non-zero effect would be at most equal to $\Delta$.  The interval $(-\Delta, \Delta)$ is known as the equivalence margin and represents a range of values for which $\theta$ is considered negligible.  The value of $\Delta$ is sometimes known as the ``smallest effect size of interest'' \citep{lakens2017equivalence}.  (Note that the equivalence margin need not necessarily be symmetric, i.e., we could have $\textrm{H}_{0}: \theta \notin [\Delta_{1}, \Delta_{2})$, where $\Delta_{1} \ne -\Delta_{2}$.)
%
%
\noindent Consider a multiple linear regression where $Y$ is the outcome variable and $X$ is the $N$ $\times$ $(K+1)$ fixed { predictor }matrix (with a column of 1s for the intercept); see  \citet{azen2009applications} for an accessible review.   Going forward, we use $X_{k}$ to refer to the $k$-th {predictor, for $k$ in 0,...,$K$}.  

Note that the regression may include both categorical and continuous predictors.  For example, suppose a researcher is looking to investigate possible predictors of anxiety among high-school students.  In this hypothetical study, $Y$ might be a student's score on an anxiety assessment questionnaire; $X_{1}$ might be a binary variable indicating whether or not the student received counselling services (0 = ``did not receive counselling; 1 = ``did receive counselling''); $X_{2}$ might be a continuous { predictor }corresponding to the student's age in years; and $X_{3}$ might be a continuous { predictor }corresponding to the student's household income in dollars.  %Table 1  provides some artificial example data to illustrate.

% \begin{table}[]
%     \centering
%     \begin{tabular}{c|c|c|c}
%          $Y$ & $X_{1}$ & $X_{2}$ & $X_{3}$\\
%          & & & &
%     \end{tabular}
%     \caption{Caption}
%     \label{tab:my_label}
% \end{table}

We operate under the standard linear regression assumption that the $N$ observations in the data are independent and normally distributed such that, for $i$=1,...,$N$:
%
%\begin{equation} Y_{i} \sim  \; \textrm{Normal}(X_{i \times}^{T}\beta, \sigma^{2}), \quad %\quad \forall \; i=1,...,N;
%\label{regress}
%\end{equation}
\begin{align*}
    Y_{i} &=\beta_{0} + \beta_{1}X_{i,1} +  \ldots + \beta_{K}X_{i,K} +\epsilon_{i}, \textrm{ 
 and}\\  
 \epsilon_{i} &\sim \textrm{Normal}(0, \sigma^{2}), 
\label{regress}
\end{align*}
\noindent {where  $\beta = (\beta_{0}, \beta_{1}, \ldots, \beta_{K})^{T}$ is a parameter vector of $K+1$ regression coefficients,  $\epsilon_{i}$ are residuals, and $\sigma^{2}$ is the population variance parameter (i.e., the variability of the random errors).}  Least squares estimates for the linear regression model are denoted by $\reallywidehat{\beta} = (\reallywidehat{\beta_{0}}, \reallywidehat{\beta_{1}}, \ldots, \reallywidehat{\beta_{K}})^{T}$, and $\hat{\sigma}^{2}$; see equations (\ref{eq:beta}) and (\ref{eq:sigma}) which are provided in the supplemental material for completeness.   

Recall that, for $k$ in $1,\ldots,K$, the interpretation of the $\beta_{k}$ coefficient is the  average change in the response variable ($Y$) for every unit change in the explanatory variable ($X_{k}$) when holding all other predictors constant.  For example, in our hypothetical study about anxiety, the $\beta_{1}$ coefficient would be interpreted as the average number of additional points on the anxiety assessment score associated with a student receiving counselling services given a fixed age and household income.
%, the $\beta_{2}$ coefficient would be interpreted as the average number of additional points on the anxiety assessment score associated with every additional year in age, and the $\beta_{3}$ coefficient would be interpreted as the average number of additional points associated with every additional dollar in household income.

An equivalence test for an unstandardized regression coefficient asks the following question: {Can we reject the possibility that $\beta_{k}$ is as large or larger than our smallest effect size of interest?}  Formally, the null and alternative hypotheses for the equivalence test are stated as:
\begin{equation}
\begin{split}
\textrm{H}_{0}&: \beta_{k} \le \Delta_{k,lower} \quad \textrm{or}  \quad \beta_{k} \ge \Delta_{k,upper}, \quad \textrm{vs.}  \\
 \indent \textrm{H}_{1}&: \beta_{k} > \Delta_{k,lower} \quad \textrm{and} \quad \beta_{k} < \Delta_{k,upper} ,
 \end{split}
 \label{eq:hyp1}
\end{equation}
 \noindent where the equivalence margin, $(\Delta_{k,lower}, \Delta_{k,upper})$, defines the range of values considered negligible, for $k$ in $0,\ldots,K$.  Often, the equivalence margin will be symmetrical such that $\Delta_{k} = \Delta_{k,upper} = - \Delta_{k,lower}$, but this is not necessarily so.  Also, in some situations, instead of a two-sided equivalence test, a one-sided equivalence test, known as a non-inferiority test, is required.  A one-sided test can defined by simply setting the margin as a one sided-interval: $(-\infty, \Delta_{k,upper})$, or as $(\Delta_{k,lower}, \infty)$; see \citet{wellek2010testing}.
 
 Returning to our hypothetical example, suppose that in order for the impact of counselling services to be considered at all meaningful, the services would have to be associated with a minimum two point difference on the anxiety assessment questionnaire.  In this case, the researcher would simply define $\Delta_{1,lower} = -2$ and $\Delta_{1,upper} = 2$.  The equivalence margin for $k=1$ would be (-2, 2).
For the other predictors, $k=2$ and $k=3$, it may be more difficult to define an equivalence margin since $\beta_{2}$ and $\beta_{3}$ are measured in terms of ``points per year'' and ``points per dollar''. To define an appropriate margin, the researcher would have to ask: {What are the minimum meaningful per year and per dollar numbers of points to consider?}

  %See \cite{counsell2015equivalence} for a review of equivalence testing procedures for unstandardized linear regression coefficients. 
 
 There is a one-to-one correspondence between an equivalence test and a confidence interval (CI); see \citet{dixon2018primer} for details.  As such, an equivalence test can be constructed by inverting a confidence interval.  For example, we will reject the above null hypothesis ($\textrm{H}_{0}: \beta_{k} \le \Delta_{k,lower} \quad \textrm{or}  \quad \beta_{k} \ge \Delta_{k,upper}$), at a $\alpha$ significance level, whenever a (1 - 2$\alpha$)\% CI for $\beta_{k}$ fits entirely within $(\Delta_{k,lower}, \Delta_{k,upper})$.  
 
 {Inverting the CI} for $\beta_{k}$ leads to two one-sided $t$-tests (TOST) with the following $p$-values:
\begin{align} \label{noninfFp}
p_{k}^{lower} & = 1-F_{t}\left(\frac{
\hat{\beta}_{k} - \Delta_{k,lower}}{{\textrm{SE}(\hat{\beta}_{k})}}, N-K-1
\right), \quad  \textrm{and} \quad \nonumber \\ \quad
p_{k}^{upper} & = 1-F_{t}\left(\frac{
\Delta_{k,upper} - \hat{\beta}_{k}}{{\textrm{SE}(\hat{\beta}_{k})}}, N-K-1
\right),
\end{align}
 \noindent for $k$ in 0,...,$K$;  where $F_{t}(\quad  \cdot \quad ; df)$ denotes the cumulative distribution function (cdf) of the $t$-distribution with $df$ degrees of freedom, and where ${\textrm{SE}({\hat{\beta}_{k}})} = \hat{\sigma}\sqrt{[(X^{T}X)^{-1}]_{kk}}$. In order to reject the equivalence test null hypothesis ($\textrm{H}_{0}: \beta_{k} \le \Delta_{k,lower} \quad \textrm{or}  \quad \beta_{k} \ge \Delta_{k,upper}$), both $p$-values, $p_{k}^{lower}$ and $p_{k}^{upper}$,  must be less than $\alpha$.  As such, for the $k$-th regression coefficient, $\beta_{k}$, a single overall $p$-value for the equivalence test can be calculated as: $p$-value$_{k}$ = $\textrm{max}(p_{k}^{lower}, p_{k}^{upper})$.
 
 {An \textit{a priori} sample size calculation for this equivalence test can be performed using the following analytic formula to obtain a reasonable approximation of the equivalence test's statistical power \citep{zhang2003simple}:}
\begin{align}
 \label{power1}
    power = F_{t}\Big(\frac{\Delta_{k,lower}-\beta_{k}}{\textrm{SE}(\hat{\beta}_{k})}-t_{1-\alpha}^{*}, N-K-1\Big)
    -F_{t}\Big(\frac{\Delta_{k,upper}-\beta_{k}}{\textrm{SE}(\hat{\beta}_{k})} +t_{1-\alpha}^{*}, N-K-1\Big),    
\end{align}
{where $t_{1-\alpha}^{*}$ is the $(1-\alpha)$th percentile of a $t$-distribution with $N-K-1$ degrees of freedom, and $\beta_{k}$ and $\textrm{SE}(\hat{\beta}_{k})$ are set to whatever values are  assumed to be true \textit{a priori}.}  {Note that if $X_{k}$ is uncorrelated with the other predictors,  ${\textrm{SE}({\hat{\beta}_{k}})} =\sigma/(\sigma_{k}\sqrt{N})$, where $\sigma_{k}$ is the standard deviation of the $k$-th predictor, $X_{k}$, for $k$ in 1,...,$K$.}  
%In the supplemental material, the ``equivBeta'' and ``equivBetaPower'' R functions are provided, corresponding to equations (\ref{noninfFp}) and (\ref{power1}), respectively. 

 {
 To exemplify the above power calculation and TOST procedure, we return to our hypothetical anxiety study example.  The parameter of primary interest in this example is $\beta_{1}$, the effect of counselling services on the anxiety score.  As noted earlier, suppose the researcher has defined  $\Delta_{1,lower} = -2$ and $\Delta_{1,upper} = 2$ and suppose that the researcher is considering collecting data from $N=40$ participants, randomly assigning half to receive counselling.  Based on what is known about the typical variability of scores obtained with the anxiety assessment questionnaire, the researcher reasonably assumes, \textit{a priori}, that $\sigma=2$. Then, to approximate the power of the equivalence test, one can calculate ${\textrm{SE}({\hat{\beta}_{1}})} =\sigma/(\sigma_{1}\sqrt{N})= 2/(0.5\times\sqrt{40})=0.63$, and, using equation (\ref{power1}), determine that the study will have about 85\% power to reject the equivalence null hypothesis ($\textrm{H}_{0}: \beta_{1} \notin (-2,2)$) if the true value of $\beta_{1}$ is in fact zero (i.e., if the counselling services truly have no effect on anxiety scores):}
 \begin{equation*}
 power = F_{t}\Big(\frac{-2-0}{0.63}-1.69, 40-3-1\Big)-F_{t}\Big(\frac{2-0}{0.63} +1.69, 40-3-1\Big)=0.85.
 \end{equation*}
 Now suppose the researcher conducts the study (see full dataset in Table \ref{tab:anx} of the supplemental material) and obtains the following results:
 \begin{align*}
   \hat{\beta}_{0} = 3.69;  \quad \textrm{SE}({\reallywidehat{\beta_{0}}}) = 3.34 ;  \quad &90\%\textrm{CI}=(-1.94,9.31); \\ \quad &95\%\textrm{CI}=(-3.08,10.45);\\
     \hat{\beta}_{1} = -0.57;  \quad {\textrm{SE}({\reallywidehat{\beta_{1}}})} = 0.66;  \quad &90\%\textrm{CI}=(-1.68,0.54); \\ \quad &95\%\textrm{CI}=(-1.90,0.76);\\
     \hat{\beta}_{2} = 0.84;  \quad {\textrm{SE}({\reallywidehat{\beta_{2}}})} = 0.20;  \quad &90\%\textrm{CI}=(0.49,1.18);\\  \quad &95\%\textrm{CI}=(0.42,1.24);\\
     \hat{\beta}_{3} = -1.93\times10^{-5};  \quad {\textrm{SE}({\reallywidehat{\beta_{3}}})} = 1.77\times10^{-5};  \quad &90\%\textrm{CI}=(-4.91,1.05)\times10^{-5}; \\ \quad &95\%\textrm{CI}=(-5.51,1.65)\times10^{-5};
 \end{align*}  
 {
These results suggest that, on average, older students obtain higher anxiety scores ($\hat{\beta}_{2} = 0.84$), and students from wealthier households  obtain lower scores ($\hat{\beta}_{3} = -1.93\times10^{-5}$; about 2 points lower for every \$100,000 increase in household income). In order to test whether $\beta_{1}$ is at most negligible, one calculates, from equation (\ref{noninfFp}):}
\begin{align*} \label{noninfFp}
p_{1}^{lower} &= 1-F_{t}\left(\frac{
\reallywidehat{\beta_{1}} - (-2)}{{\textrm{SE}(\reallywidehat{\beta_{1}})}}, N-K-1
\right) = 1-F_{t}\left(2.16, 36\right)=0.0178 , \quad  \textrm{and} \quad \\ \quad
p_{1}^{upper} &= 1-F_{t}\left(\frac{
2 - \reallywidehat{\beta_{1}}}{{\textrm{SE}(\reallywidehat{\beta_{1}})}}, N-K-1
\right)  = 1-F_{t}\left(3.92, 36\right)=0.0002,
\end{align*}
such that:
\begin{align*} 
\quad p-\textrm{value}_{1} &= \textrm{max}(p_{1}^{lower}, p_{1}^{upper}) = 0.0178.
\end{align*}
 {
 If a nominal significance level of $\alpha=0.05$ is to be used for the equivalence test, then we can reject the equivalence null hypothesis ($\textrm{H}_{0}: \beta_{1} \notin (-2,2)$) and  plausibly conclude that, when controlling for age and wealth, any difference on the anxiety score between those students receiving counselling and those not receiving counseling is smaller than 2 points ($p$-value=0.0178).}
{\subsection*{The ``least equivalent allowable difference'' or ``equivalence confidence interval''}}
{\citet{seaman1998equivalence} and \citet{meyners2007least} suggest that researchers, instead of only reporting an equivalence test $p$-value, should also report the smallest possible absolute value at which one could have claimed equivalence.  \citet{meyners2007least} calls this the ``least equivalent allowable difference'' (LEAD), while \citet{seaman1998equivalence} refer to this as the ``equivalence confidence interval.''  To illustrate the concept, we return one again to our hypothetical anxiety study.}

{Instead of simply concluding that any difference on the anxiety score between those students receiving counselling and those not receiving counseling is smaller than 2 points with $p$-value = 0.0178, one could also report that any difference less than 1.68 could also have been ruled out (at the nominal significance level of $\alpha=0.05$).  Indeed, prior to having observed the data, had we defined the equivalence margin to be (-1.68, 1.68) instead of (-2, 2),  then we could have rejected the equivalence null hypothesis with $p$-value=0.05.  In this case, the LEAD is equal to 1.68 and the ``equivalence confidence interval'' is simply (-1.68,1.68).  }%This interval corresponds to the widest possible symmetric equivalence margin that one could have hypothetically defined (prior to having observed the data), for which one can reject the null hypothesis of non-equivalence. }

{Note that the 90\% CI for $\beta_{1}$ is (-1.68, 0.54).  The -1.68 lower bound is no coincidence.  The LEAD can be calculated as the maximum of the absolute value of the bounds of the $(1-2\alpha)\%$CI.  For instance, in our example, we have that LEAD$=\textrm{max}(|-1.68|,|0.54|) = 1.68$.  Despite it's simplicity, \citet{meyners2007least} argues that the LEAD is worth reporting since it depends only on the $\alpha$ significance level and thereby enables readers to draw their own conclusions irrespective of the equivalence margin that a particular researcher might choose.}
\section{Equivalence testing for a standardized effect size in linear regression}
\label{sec:stdbeta}
%
%As discussed in the Introduction, in many situations, rather than conduct an equivalence test for an unstandardized regression coefficient, it may be preferable to work with standardized regression coefficients.  
%
Unstandardized regression coefficients are often difficult to interpret since both the predictors and the outcome can be measured on arbitrary units with no objective meaning.  As a result, researchers may prefer to report standardized effect sizes.  Unfortunately, equivalence testing with standardized effects is not always straightforward.  Contrary to certain recommendations, one cannot merely define the equivalence margin in terms of a standardized effect size and proceed as normal.  For example, \cite{lakens2017equivalence}'s suggestion that, for a two-sample test for the equivalence in means, one may simply define the equivalence margin in terms of the observed standard deviation is incorrect.  

The equivalence margin cannot be defined as a function of the observed data as this will invalidate the test.  Instead, one must define the parameter of interest to be the standardized parameter, such that the randomness associated with standardization is properly taken into account.  To explain why, let us consider a two-sample equivalence test for the difference in means.

Suppose that, for the difference in means, $\mu_{d}$, one were to define a symmetric equivalence margin, $(-\Delta, \Delta)$, in terms of the observed standard deviation,  $\hat{\sigma}$, such that $\Delta = 0.5\times\hat{\sigma}$.  \cite{lakens2018equivalence} consider this example and claim (incorrectly) that ``when the equivalence bounds are based on standardized differences, the equivalence test depends on the standard deviation in the sample.''   Recall that in order for a hypothesis test to be valid, the hypotheses must be statements about the unobserved parameters and not about the observed sample.  Therefore, since the hypotheses for the test in the example, $\textrm{H}_{0}: |\mu_{d}| \ge 0.5\times\hat{\sigma}$, vs. $\textrm{H}_{1}: |\mu_{d}| < 0.5\times\hat{\sigma}$, are defined as functions of the observed data (i.e., in terms of $\hat{\sigma}$), the test is invalid.  
 
 Instead, the correct procedure is to  define the parameter of interest, $\theta$, to be the standardized effect size, e.g. define $\theta= \mu_{d}/\sigma$.  Then, one can define the margin on the standardized scale without invalidating the hypotheses.  To be clear,   $\textrm{H}_{0}: |\theta| \ge 0.5$ vs.  $\textrm{H}_{1}: |\theta| < 0.5$ is a completely valid test, while $\textrm{H}_{0}: |\mu_{d}| \ge 0.5\times\hat{\sigma}$, vs. $\textrm{H}_{1}: |\mu_{d}| < 0.5\times\hat{\sigma}$ is invalid.  In this example, the valid equivalence test requires the use of a non-central $t$-distribution; see supplemental material for details on how to conduct the valid test and \citet{weber2012testing} for a worked-through example.

 While in practice, the difference between setting $\textrm{H}_{0}: |\theta| \ge 0.5$ and $\textrm{H}_{0}: |\mu_{d}| \ge 0.5\times\hat{\sigma}$ may be small, it should nevertheless be acknowledged since one should always (ideally) take into account the uncertainty involved in estimating the standard deviation.  In the supplemental material, we show results from a small simulation study (Simulation Study 1) which suggest that, in practice, using the invalid test can lead to a higher than advertised type 1 error when sample sizes are large, and a minor loss of efficiency when sample sizes are small.  {This is likely the result of failing to account for the uncertainty involved in estimating the standard deviation.}

{The most commonly used standardized effect sizes for linear regression analyses are the standardized regression coefficient and the semipartial correlation coefficient \citep{courville2001use, dudgeon2016comparative}.  However, as \citet{dudgeon2016comparative} notes, the popularity of the standardized regression coefficient ``is arguably a product of convention rather than any perceived intrinsic merit of the standardized regression coefficient as an effect size.''  Indeed, many researchers argue that the standardized regression coefficient is difficult to interpret (and problematic when it comes comparing effect sizes across different studies) since it does not appropriately partition variance when predictors are correlated \citep{ kanetkar1995effect,tonidandel2011relative,aloe2012effect}.  \citet{levine2008communication} explain as follows: ``the common practice of interpreting [the standardized regression coefficient] as analogous to [the correlation] [...] can be misleading because [the standardized regression coefficient] can be near zero even when the predictor explains a substantial amount of the variance in the outcome variable when other predictors correlated with the predictor claim the shared variance.''  For a more detailed explanation see \citet{disbato2016}  and \citet{darlington2017regression} who argue that the semipartial correlation coefficient is a much better effect size.  {Unlike the standardized regression coefficient, the semipartial correlation coefficient will always be commensurate with the amount of variance in the outcome variable that is explained by the predictor.}  With this in mind, we propose that researchers use an equivalence test for the semipartial correlation coefficient.}
\subsection*{An equivalence test for the semipartial correlation coefficient}
{The semipartial correlation coefficient, $sr_{k}$, is a parameter taking values between -1 and 1, that measures the strength of the association between the outcome, $Y$, and the predictor, $X_{k}$, that is independent of any linear relationship between $X_{k}$ and the other predictors in the model.  We define $sr_{k}$ as follows, for $k$ in 1,...,$K$:}
\begin{equation}
sr_{k} =  \Big(\beta_{k} \frac{\sigma_{k}}{\sigma_{Y}}\Big)\times  \sqrt{1-R^{2}_{X_{k}X_{-k}}},    
\end{equation}
{\noindent where $\sigma_{Y}$ is the standard deviation of $Y$, $\sigma_{k}$ is the standard deviation of $X_{k}$, and $R^{2}_{X_{k}X_{-k}}$ is the coefficient of determination from the linear regression of $X_{k}$ predicted from $X_{-k}$, all of the other $K-1$ predictors (see full details in supplemental material).  Note that $\mathcal{B}_{k} = (\beta_{k} {\sigma_{k}}/{\sigma_{Y}})$ is equal to the $k$-th standardized regression coefficient, and that ${1-R^{2}_{X_{k}X_{-k}}}$ is known as the $k$-th ``tolerance'', representing the proportion of variation in $X_{k}$ that is linearly unrelated to all the other predictors in the regression model.  To be clear, when $K=1$ (or when $X_{k}$ and $X_{-k}$ are perfectly uncorrelated),  the tolerance will equal 1, and therefore: $sr_{k} = \mathcal{B}_{k} = \textrm{cor}(Y, X_{k})$.  }  
%the semipartial correlation coefficient will therefore equal the standardized regression coefficient, and will also be equal to the bivariate correlation between $Y$ and $X_{k}$ such that

{Note that the \textit{squared} semipartial correlation, $sr_{k}^{2}$, can be understood as the amount of variance in $Y$ that is uniquely explained by the $k$-th predictor since:}
\begin{equation}
\label{eq:sr}
    sr_{k}^{2}= \Big(\beta_{k} \frac{\sigma_{k}}{\sigma_{Y}}\Big)^{2}\times  \Big({1-R^{2}_{X_{k}X_{-k}}}\Big) = R^{2}_ {Y X} - R^{2}_{Y X_{-k}},
\end{equation}
{\noindent where $R^{2}_ {Y X}$ is the coefficient of determination from the linear regression of $Y$ predicted from $X$, and where $R^{2}_{Y X_{-k}}$ is the coefficient of determination from the linear regression of $Y$  predicted from all but the $k$-th predictor.}   

{\citet{cohen1988statistical}'s well known rules of thumb for interpreting the magnitude of correlation coefficients (small=0.1, medium=0.3, large=0.5) can be applied for interpreting semipartial correlation coefficients.  However, we note that \citet{cohen2013applied} offer alternative values that are slightly larger: small effects may be defined as $sr_{k}=0.14$ (or equivalently $sr_{k}^{2}=0.02$), medium effects as $sr_{k}=0.39$ (or $sr_{k}^{2}=0.15$), and large effects as $sr_{k}=0.59$ (or $sr_{k}^{2}=0.35$).  Determining what exact values are ideal is beyond the scope of this paper, but interested readers are referred to \citet{hemphill2003interpreting}, \citet{funder2019evaluating}, and \citet{lovakov2021empirically}.}

{\citet{dudgeon2016comparative} propose using the adjusted Aloe-Becker large-sample confidence interval for $sr_{k}$ which can be  calculated, for $k$ in 1,...,$K$, as:}
\begin{equation}
(1-\alpha)\%\textrm{CI} \textrm{ for } sr_{k} = [\reallywidehat{sr_{k}}-t^{*}_{1-\alpha/2}\textrm{SE}(\reallywidehat{sr_{k}}), \quad \reallywidehat{sr_{k}}+t^{*}_{1-\alpha/2}\textrm{SE}(\reallywidehat{sr_{k}})],
\end{equation}
\noindent {where:}
\begin{equation}
\label{eq:sr_hat}
\reallywidehat{sr_{k}} =  (\hat{\beta}_{k} \frac{\hat{\sigma}_{k}}{\hat{\sigma}_{Y}})\times  \sqrt{1-\hat{R}^{2}_{X_{k}X_{-k}}},    
\end{equation}
{and}
\begin{equation}
\label{eq:SEsr_hat}
\textrm{SE}(\reallywidehat{sr_{k}}) = \sqrt{
\frac{\hat{R}_{YX}^4 - 2\hat{R}_{YX}^2 + \hat{R}_{YX_{-k}}^2 + 1 - \hat{R}_{YX_{-k}}^4}{N-K-1}
},
%     \label{eq:SEbetastd_hat}
\end{equation}
{where $\hat{R}_{YX}^{2}$, $\hat{R}_{YX_{-k}}^2$,  and $\hat{R}_{X_{k}X_{-k}}^2$ are estimates for ${R}_{YX}^{2}$, ${R}_{YX_{-k}}^2$, and ${R}_{X_{k}X_{-k}}^2$, respectively, obtained from the observed data.} 

{An equivalence test for the $k$-th semipartial correlation coefficient can be defined by the following null and alternative hypotheses:}

$\textrm{H}_{0}: sr_{k} \le \Delta_{k,lower} \quad \textrm{or:}  \quad sr_{k} \ge \Delta_{k,upper} $, \quad \textrm{vs.}\\
 \indent $\textrm{H}_{1}: sr_{k} > \Delta_{k,lower} \quad \textrm{and:} \quad sr_{k} < \Delta_{k,upper}, $

\noindent {where the equivalence margin is $(\Delta_{k,lower}, \Delta_{k,upper})$, for $k$ in $1,\ldots,K$.  By inverting the adjusted Aloe-Becker large-sample confidence interval, we can conduct two one-sided $t$-tests (TOST) with the following $p$-values, for $k$ in 1,...,$K$:}
\begin{align}
\mathbbmss{p}_{k}^{lower} &= 1-F_{t}\left(
\frac{\reallywidehat{sr}_{k}-\Delta_{k,lower}}{{\textrm{SE}(\reallywidehat{sr}_{k})}}; df = N-K-1
\right), \quad \nonumber \textrm{and} \\ 
\mathbbmss{p}_{k}^{upper} &= 1-F_{t}\left(
\frac{\Delta_{k,upper}-\reallywidehat{sr}_{k}}{{\textrm{SE}(\reallywidehat{sr}_{k})}}; df = N-K-1
\right). \label{eq:spcpval}
\end{align}
\noindent {Therefore, for the $k$-th predictor, the null hypothesis ($\textrm{H}_{0}: sr_{k} \le \Delta_{k,lower} \quad \textrm{or:}  \quad sr_{k} \ge \Delta_{k,upper}$) is rejected if and only if  ${p}$-value$_{k}$ = $\textrm{max}(\mathbbmss{p}_{k}^{lower}, \mathbbmss{p}_{k}^{upper})$, is less than $\alpha$.  }

{An \textit{a priori} sample size calculation for this equivalence test can be performed using the following analytic formula to obtain a reasonable approximation of the equivalence test's statistical power:}
\begin{equation}
 \label{power2}
    power = F_{t}\Big(\frac{\Delta_{k,lower}-sr_{k}}{\textrm{SE}(\reallywidehat{sr}_{k})}-t^{*}_{1-\alpha}, N-K-1\Big)-F_{t}\Big(\frac{\Delta_{k,upper}-sr_{k}}{\textrm{SE}(\reallywidehat{sr}_{k})} +t^{*}_{1-\alpha}, N-K-1\Big),
\end{equation}
{where  $sr_{k}$ and ${\textrm{SE}({\reallywidehat{sr}_{k}})}$ are whatever values are assumed to be true \textit{a priori}.  Note that if one assumes that $sr_{k}=0$, then ${\textrm{SE}(\reallywidehat{sr}_{k})}=\sqrt{1/(N-K-1)}$.  }

{In the supplemental material, we conduct two small simulation studies, Simulation Study 2 and Simulation Study 3, to investigate the proposed methods.  The first shows that, when $K=1$, the proposed equivalence test for semipartial correlation coefficients and a commonly used equivalence test for correlations (based on Fisher's $Z$ transformation) provide, as expected, very similar results.  The second simulation study confirms that the type 1 error obtained with the proposed equivalence test for semipartial correlation coefficients is correct and that the proposed formula for power calculation (equation (\ref{power2})) provides a reasonable approximation of the true statistical power.  The results also suggest that large sample sizes (much larger than those typically encountered in psychological studies \citep{kuhberger2014publication, fraley2014n, marszalek2011sample}) may be required for the equivalence test to have non-negligible statistical power.  \citet{goertzen2010detecting} reached a similar conclusion.}

\section{A Bayesian alternative for establishing equivalence in a linear regression}
\label{sec:bayes}

%\subsection{Bayes Factor testing for linear regression}

As noted in the Introduction, there are a several different Bayesian methods available for establishing equivalence.  \cite{rouder2012default}'s proposed ``default'' Bayes factor (based on the work of \cite{liang2008mixtures}) is one approach that has proven to be particularly popular in psychology research for linear regression models \citep{etz2015using, morey2015package}.  We briefly review the default Bayes factor approach for linear regression in order to consider how it might compare to the frequentist equivalence tests we proposed.

The Bayes Factor, $\textrm{BF}_{10}$, is defined as the probability of the data under the alternative model relative to the probability of the data under the null model:
\begin{equation}
\textrm{BF}_{10} = \frac{\textrm{Pr}(Data\,|\,Model\:{1})}{\textrm{Pr}(Data\,|\,Model\:{0})} = \frac{\textrm{Pr}(\,Model\:{1}|Data\,) \times \textrm{Pr}(\,Model\:{1})}{\textrm{Pr}(\,Model\:{0}|Data\,) \times \textrm{Pr}(\,Model\:{0})},
\end{equation}
\noindent with the ``10'' subscript indicating that the alternative model (i.e., ``Model 1'') is being compared to the null model (i.e., ``Model 0'').  {Interpretation of the Bayes factor is straightforward.  For example, with equal prior model probabilities, a $\textrm{BF}_{10}$ equal to $0.20$ indicates that the null model is five times more likely than the alternative model.  Going forward, we suppose that equal prior model probabilities ($\textrm{Pr}(\textrm{Model 0}) = \textrm{Pr}(\textrm{Model 1}) = 0.5$) are always assumed, as is often (implicitly) done in practice; but see \citet{tendeiro2019review}, and \citet{campbell2022coin} for discussion of this practice.}

{Bayesian methods require one to define appropriate prior distributions for all model parameters \citep{consonni2008compatibility} and  \cite{rouder2012default} suggest using  Jeffreys-Zellner-Siow (JZS) ``objective priors''.  A version of this prior setup, whereby the so-called $r$ scale parameter is set equal to a specific value, allows one to specify prior beliefs about the magnitude of standardized regression coefficients (i.e., specify the \textit{a priori} 
 distribution of $\mathcal{B}_{k} = (\beta_{k} {\sigma_{k}}/{\sigma_{Y}})$, for $k$ in 1,...,$K$).  For instance, the BayesFactor R package uses the scaled-JZS prior setup with a default of $r = {\sqrt{2}}/{4}= 0.354$, corresponding to a prior belief in a 50\% probability that $|\mathcal{B}_{k}|> 0.354$, for $k$ in 1,...,$K$.  While perhaps ``computationally convenient'', researchers who have difficulty interpreting the magnitude of standardized regression coefficients (e.g., \citet{disbato2016}) will no doubt be particularly challenged when it comes to defining an appropriate value for $r$.}

 %The Bayes factor {outlined above} is scale invariant so that it will not change if the regression coefficients are measured in different units.  However, since the priors are set to the standardized regression coefficients, the Bayes factor would change if the correlation between covariates were to change. ZZZ

%Note that, just like frequentist testing of semipartial correlation coefficients, Bayes factor testing {outlined above} is scale invariant so that it is entirely independent of the specific measurement units  (i.e., the Bayes factor will not change if the regression coefficients are measured in different units).  However,   

To test the $k$-th regression coefficient, in a multiple linear regression model, one computes a Bayes factor for a model that includes the $k$-th { predictor }against a model that does not, such that, for  $i$=1,...,$N$:
\begin{align}\textrm{Model }\, 0 &: \quad Y_{i}    \sim   \textrm{Normal}(X_{i,-k}^{T}\beta_{-k}, \sigma^{2}), \quad \quad   \textrm{and} \nonumber \\
\textrm{Model }\, 1 &: \quad Y_{i}   \sim   \textrm{Normal}(X_{i \times}^{T}\beta, \sigma^{2}), \quad \quad \nonumber
\label{models}
\end{align}
\noindent {where $\beta_{-k}$ is the vector of regression coefficients with the $k$-th coefficient  omitted,  $X_{i \times}$ represents all $K+1$ predictor values corresponding to the $i$-th observation, and $X_{i,-k}$ is simply  $X_{i \times}$ with the $k$-th predictor value omitted.} 

If this Bayes factor were to be above a certain threshold (e.g., if $\textrm{BF}_{10} > 6$), one would conclude with a ``positive'' finding that $\beta_{k}$ is different than 0 (i.e., evidence in support of Model 1).  On the other hand, if this Bayes factor were to be bellow a certain threshold (e.g., if $\textrm{BF}_{10} < 1/6$), one would conclude with a ``negative'' finding that there is evidence for $\beta_{k}=0$ (i.e., evidence in support of Model 0).   Finally, if this Bayes factor were neither above or bellow the certain threshold (e.g., if $1/6 < \textrm{BF}_{10} < 6$), one would conclude with a ``inconclusive'' finding that there is insufficient evidence to support either model.  

{\cite{campbell2018conditional} discuss a similar frequentist way to categorize one's results as either ``positive'', ``negative'', or ``inconclusive''.  Testing a parameter $\theta$ under the so-called ``conditional equivalence testing'' (CET) scheme would proceed as follows.  If a first $p$-value, $p_{NHST}$, obtained from testing $\textrm{H}_{0}:\theta=0$, is less than the type 1 error $\alpha$-threshold (e.g., if $p_{NHST} < 0.05$), one concludes with a ``positive'' finding: $\theta$ is significantly different than 0.  On the other hand, if the first $p$-value, $p_{NHST}$, is greater than $\alpha$ and a second $p$-value, $p_{EQUIV}$, obtained from testing $\textrm{H}_{0}:|\theta|>\Delta$, is smaller than $\alpha$ (e.g., if $p_{NHST} \ge 0.05$ and $p_{EQUIV} < 0.05$), one concludes with a ``negative'' finding: there is evidence of statistically significant equivalence.  Finally, if both $p$-values are larger than $\alpha$, the result is ``inconclusive''.  In the next section we will have the opportunity to see how the BF and CET based categorizations compare.}

\section{Practical Examples}

\subsection*{Evidence for gender bias -or the lack thereof- in academic salaries}\label{sec:app}

As a first example to illustrate the various testing methods, we turn to the ``Salaries'' dataset (from R CRAN package “car”; see \citet{fox2012package}).  This dataset has been used as an example in other work: as an example for ``anti-NHST'' statistical inference in \citet{briggs2019replacement}; and as an example for data visualization methods in \citet{moon2017learn} and \citet{ghashimggplot2}.

The data consist of a sample of salaries of university professors collected during the 2008-2009 academic year.  In addition to the posted salaries (a continuous variable, in \$US), the data includes 5 additional variables of interest: (1) sex (2 categories: (1) Female, (2) Male);
(2) years since Ph.D. (continuous, in years);
(3) years of service (continuous, in years);
(4) discipline  (2 categories: (1) theoretical, (2) applied).
(5) academic rank (3 categories: (1) Asst. Prof. , (2) Assoc. Prof., (3) Prof.).

The sample includes a total of $N=397$ observations with 358 observations from male professors and 39 observations from female professors. The minimum measured salary is \$57,800, the maximum is \$231,545, and the median salary is \$107,300.   A primary question of interest is whether there is a difference between the salary of a female professor and a male professor when accounting for possible observed confounders: rank, years since Ph.D., years of service, and discipline.  The mean salary for male professors in the sample is \$115,090, while the mean  salary for female professors in the sample is \$101,002.  For illustration purposes, we consider both a simple linear regression ($K=1$) (ignoring the confounders) and a multiple linear regression ($K=6$).
 
 \subsubsection*{A simple linear regression}
 
Consider a simple linear regression (i.e., $Y \sim \textrm{Normal}(\beta_{0} + \beta_{1}X_{1}, \sigma^{2})$) for the association between salary ($Y$, measured in \$) and sex ($X_{1}$, where ``0'' corresponds to ``female,'' and ``1'' corresponds to ``male.'').   Standard least squares estimation results in the following parameter estimates:
      $\reallywidehat{\beta}_{0} = 101002$, $\operatorname{SE}(\reallywidehat{\beta}_{0}) = 4809$, and $\reallywidehat{\beta}_{1} = 14088$, $\operatorname{SE}(\reallywidehat{\beta}_{1}) = 5065$;
 $\reallywidehat{\sigma} =  30034.61$; $\reallywidehat{sr}_{1} = 0.14$, $\operatorname{SE}(\reallywidehat{sr}_{1})=0.05$.

  We can conduct an equivalence test to determine if the difference in salaries between male and female professors is at most no more than some negligible amount.  Suppose that any difference of less than $\Delta=\$5,000$ is considered negligible.  Then a $p$-value for the equivalence test, $\textrm{H}_{0}: |{\beta}_{1}| \ge 5000 \textrm{ vs. } \textrm{H}_{1}: |{\beta}_{1}| < 5000$, can be calculated following equation (\ref{noninfFp}).  We obtain  $p-\textrm{value}_{1}$ = $\textrm{max}(p_{1}^{lower},p_{1}^{upper}) =\textrm{max}(0.00009,0.963) =0.963$ and therefore fail to reject the equivalence test null hypothesis.  
  
  {If it were not possible to determine a specific number of dollars to be considered negligible, we could conduct an equivalence test for the semipartial correlation coefficient, ${sr}_{1}$.  Suppose we consider anything less than ``small'' to be negligible and therefore define the equivalence margin as (-0.10, 0.10).  Then we can calculate a $p$-value for $\textrm{H}_{0}: |{sr}_{1}| \ge 0.10 \textrm{ vs. } \textrm{H}_{1}: |{sr}_{1}| < 0.10 $  as per equation (\ref{eq:spcpval}).  We obtain:}
 %    
%$\quad$
%
 $p\textrm{-value}=\textrm{max}(\mathbbmss{p}_{1}^{lower}, \mathbbmss{p}_{1}^{upper})=0.783$,  where:
\begin{align*}
\mathbbmss{p}_{k}^{lower} &= 1-F_{t}\left(
\frac{\reallywidehat{sr}_{k}-\Delta_{k,lower}}{{\textrm{SE}(\reallywidehat{sr_{k}})}}; df = N-K-1
\right), \\
&= 1-F_{t}\left(\frac{0.14+0.10}{0.05}; df = 397-1-1 \right),  \\ 
&= 1-F_{t}\left(2.78; df = 395 \right),  \\ 
&< 0.001 
\end{align*}
\noindent and:
\begin{align*}
\mathbbmss{p}_{k}^{upper} &= 1-F_{t}\left(
\frac{\Delta_{k,upper}-\reallywidehat{sr}_{k}}{{\textrm{SE}(\reallywidehat{sr}_{k})}}; df = N-K-1\right), \\
&= 1-F_{t}\left(\frac{0.10-0.14}{0.05}; df = 397-1-1 \right),  \\ 
&= 1-F_{t}\left(-2.78; df = 395 \right),  \\ 
&= 0.783
\end{align*}
   Bayes factors are easy to compute as well.  With the BayesFactor package and the ``regressionBF'' function (with the default prior-scale $r=\sqrt{2}/4$), we obtain  $\textrm{BF}_{10}= 4.5$ which suggests that the alternative model (i.e., the model with ``sex'' included) is about four and a half times more likely than the null model (i.e., the intercept only model).  Note that we obtain the identical result using the ``linearReg.R2stat'' function.  However, when using the ``lmBF'' function, we obtain a value of $\textrm{BF}_{10}=6.2$ which suggests that the alternative model is about 6 times more likely than the null model.  Both functions are comparing the two very same models so this result is somewhat surprising.\footnote{The apparent contradiction can be explained by the fact that the two ``default BF'' functions are using different ``default priors.''  The ``regressionBF'' function (as we are using it, see supplemental material) assumes ``sex'' is a continuous variable, while the ``lmBF'' function assumes that ``sex'' is a categorical variable.  The ``default priors'' are defined accordingly, in different ways.  This may strike one as rather odd, since both models are numerically identical.  However, others see logic in such practice:  \citet{rouder2012default} suggest, somewhat vaguely, that researchers ``be mindful of some differences when considering categorical and continuous covariates'' and ``recommend that researchers choose priors based on whether the covariate is categorical or continuous''; see Section 13 of \citet{rouder2012default} for details.}
  
\subsubsection*{Multiple linear regression}  
  
Now consider a multiple linear regression model, with $K=6$:
 \begin{equation*}
     Y \sim \textrm{Normal}(\beta_{0} + \beta_{1}X_{1} + \beta_{2}X_{2} + \beta_{3}X_{3} + \beta_{4}X_{4} + \beta_{5}X_{5} + \beta_{6}X_{6}, \sigma^{2}),
 \end{equation*}
\noindent where $X_{1}=0$ corresponds to ``female,'' and $X_{1}=1$ corresponds to ``male'';  $X_{2}$ corresponds to years since Ph.D.;  $X_{3}$ corresponds to years of service; $X_{4}=0$ corresponds to ``theoretical,'' and $X_{4}=1$ corresponds to ``applied''; and where ($X_{5}=0$, $X_{6}=0$) corresponds to ``Asst. Prof.'', ($X_{5}=1$, $X_{6}=0$) corresponds to ``Assoc. Prof.'', and ($X_{5}=0$, $X_{6}=1$) corresponds to ``Prof.''.  

{Table \ref{tab:salaries1} lists parameter estimates obtained by standard least squares estimation as well as LEAD values for the semipartial correlation coefficients (with $\alpha=0.05$).  Table \ref{tab:salaries2} lists the $p$-values for each of the hypothesis tests we consider as well as Bayes factors.  An equivalence margin of (-0.10, 0.10) is used for the equivalence testing of the semipartial correlation coefficients and the Bayes factors are calculated using the ``regressionBF'' function from the BayesFactor package (with the default prior-scale $r=\sqrt{2}/4$).}

We obtain a Bayes factor for $k=1$ of $B_{10}= 1/3.9$, indicating only moderate evidence in favour of the null model.  This corresponds to an ``inconclusive'' result with a Bayes factor threshold of 6, or 10 (or any threshold higher than 3.9 for that matter).  The result for $k=1$ from CET would also be ``inconclusive'' (for $\alpha=0.05$ and $\Delta=0.10$), since both the NHST $p$-value ($=0.216$) and the equivalence test $p$-value ($=0.076$) are larger than $\alpha=0.05$.  As such, we conclude that, when controlling for observed confounders, there are insufficient data to support either an association, or the lack of an association, between sex and salary.  More data will be required to answer the question.  This inconclusive result might motivate researchers to undertake another study on the question with a much larger sample size.

{Note that the conclusions obtained with the CET and Bayes factor approaches do not entirely agree for the other predictors, see Table \ref{tab:salaries2}.  For both the ``years since Ph.D" ($k=2$) and the ``years of service'' ($k=3$) predictors, the frequentist CET obtains a positive result whereas the Bayes factor obtains an inconclusive result.}
 %
 % latex table generated in R 3.6.0 by xtable 1.8-4 package
% Thu Mar 26 23:56:00 2020
\begin{table}[h!!!]
\centering
\begin{tabular}{rrrrrrr}
  \hline
$k$ & { predictor }&$\beta_{k}$ & $\textrm{SE}(\hat{\beta}_{k})$ & ${\reallywidehat{sr}_{k}}$ & ${\textrm{SE}(\reallywidehat{sr}_{k})}$ & {${\textrm{LEAD}(\reallywidehat{sr}_{k})}$}  \\ 
  \hline
0 & intercept & 65955.23 & 4588.60 & -  & - & -  \\
1 & sex (male) & 4783.49 & 3858.67 & 0.046 & 0.037  & 0.108 \\  
2 & years since Ph.D. & 535.06 & 240.99 & 0.083 & 0.037 & 0.145 \\  
3 & years of service & -489.52 & 211.94 & -0.086 & 0.037 & 0.148 \\ 
4 & discipline (applied) &14417.63 & 2342.88 & 0.230 & 0.037 & 0.291 \\ 
5 & rank (Asst. Prof.) &  12907.59 & 4145.28 & 0.116 & 0.037 & 0.178 \\  
6 &  rank (Prof.)& 45066.00 & 4237.52 & 0.398 & 0.036 & 0.457 \\ 
\hline
& & && $\hat{\sigma}=22538.65$& &$R^{2}_{Y,X}=0.455$ \\
\hline
\end{tabular}
\caption{Parameter estimates obtained by standard least squares estimation for the full multiple linear regression model.}
\label{tab:salaries1}
\end{table}
 \begin{table}[hbt!]
\centering
\begin{tabular}{rrrrrrr}
  \hline
$k$  & ${\reallywidehat{sr}_{k}}$ & $p_{NHST}$ & $p_{EQUIV}$ & $\textrm{BF}_{10}$ & CET conclusion & Bayesian conclusion\\ 
&&&$\Delta=0.10$ & $r = {\sqrt{2}}/{4}$ &  $ \alpha=0.05$ & BF threshold = 6 \\
  \hline
1  & 0.046 & 0.216 & 0.076 & 1/3.9 &  Inconclusive & Inconclusive \\ 
2   &  0.083 & 0.027 & 0.325 & 1.4  & Positive & Inconclusive \\ 
3   & -0.086  & 0.021 & 0.358 & 1.7  & Positive & Inconclusive \\
4   &  0.230  & $<0.001$ & $1.000$ & $6.5\times10^{6}$ & Positive & Positive \\  
5   & 0.116 & 0.002 & 0.670 & 13.6  & Positive & Positive \\  
6  & 0.398 & $<0.001$ & $1.000$ & $1.8\times10^{20}$ & Positive & Positive\\ 
\hline
\end{tabular}
\caption{Calculated values and conclusions for both frequentist and Bayesian testing for the salaries multiple linear regression model.}
\label{tab:salaries2}
\end{table}
  \subsection*{Six key premises of mindset theory}
 As a second practical example, we consider \citet{burgoyne2020firm} who obtained data from $N=438$ individuals and fit several simple linear regressions to these data in order to investigate six key premises of ``mindset theory.''  For each of the six key premises, \citet{burgoyne2020firm} regressed a different continuous variable against an individual's ``mindset score'' and used a non-inferiority test (i.e., a one-sided equivalence test) to determine whether the correlation was significantly smaller, or larger, than a predetermined value.  Specifically, \citet{burgoyne2020firm} defined the non-inferiority margin as either -0.2 or 0.2 (depending on the direction of the effect predicted by mindset theory).  {\citet{burgoyne2020firm} justify this choice of margin by citing \citet{richard2003one} and  explaining that ``effects described as profound should at least meet the mean effect size in social-psychological research.''}
 
\citet{burgoyne2020firm} used the test for correlations proposed by \citet{goertzen2010detecting} based on Fisher's $Z$ transformation (see details of this test in supplemental material).   We calculated $p$-values for each of the six regressions based instead on our proposed test for semipartial correlation coefficients (recall that when $K=1$, $sr_{k} = \mathcal{B}_{k} = \textrm{cor}(Y, X_{k})$).  In Table \ref{tab:goertsen}, the $p$-values calculated based on equation (\ref{eq:spcpval}) are listed alongside the $p$-values obtained by \citet{burgoyne2020firm}.  We note that for each of the six simple linear regressions, the two $p$-values are very similar.
 
  \citet{burgoyne2020firm} also wished to investigate whether the association between the ``Raven failure score'' and the mindset score is no more than negligible when controlling for cognitive ability.  This requires a multiple linear regression and \citet{goertzen2010detecting}'s test for correlations is therefore not applicable.  Our proposed test for semipartial correlation coefficients is, on the other hand, well-suited for the task.  In row 7 of Table \ref{tab:goertsen}, we list the $p$-value obtained using equation (\ref{eq:spcpval}) as $p_{sr}<0.001$.
\begin{table}[h!]
\centering
\begin{tabular}{lllrrr}
  \hline
 & $\textrm{H}_{0}$ &$Y \sim X$  & {$\reallywidehat{sr}_{1}$}  & $p_{Z}$ & {$p_{sr}$} \\ 
  \hline
1.&  $sr_{1} \ge 0.2$ & Learning goals  & &  & \\ 
& & $\quad \sim$ Mindset & 0.098 & 0.015 & 0.016 \\ 
  \hline
2.& $sr_{1} \le -0.2$   &   Performance goals &  &  &  \\ 
& &$\quad \sim$ Mindset & -0.109 & 0.026 & 0.027 \\ 
  \hline
3. & $sr_{1} \le -0.2$  &    Performance avoidance goals&   & &  \\ 
&& $\quad \sim$ Mindset &  -0.039 & $<$0.001 & $<$0.001 \\ 
  \hline
4.  & $sr_{1} \le -0.2$  &   Belief in talent alone &  &  &  \\ 
& &$\quad \sim$ Mindset &  -0.061 & 0.002 & 0.002 \\ 
  \hline
5. &   $sr_{1} \ge 0.2$&   Response to challenge &  &  &  \\ 
& &$\quad \sim$ Mindset & 0.056 & 0.001 & 0.001 \\ 
  \hline
6. &   $sr_{1} \ge 0.2$&   Raven failure score &  &  &  \\ 
& &$\quad \sim$ Mindset & -0.122 & $<$0.001 & $<$0.001 \\ 
  \hline
7.&    $sr_{1} \ge 0.2$&   Raven failure score &  &  &  \\ 
& & $\quad \sim$ Mindset + & -0.055 & \textemdash & $<$0.001 \\ 
& & $\quad \quad \quad$ Cognitive ability   &  &  &  \\ 
   \hline
\end{tabular}
\caption{For each of the regression analyses fit by \citet{burgoyne2020firm},  $p_{sr}$ indicates the $p$-value for the equivalence test based on equation (\ref{eq:spcpval}), and $p_{Z}$ indicates the $p$-value for the equivalence test based on Fisher’s Z transformation.  Note that in order to conduct a non-inferiority test (a one-sided equivalence test), one defines an open ended equivalence margin (which we indicate by setting either $\Delta_{lower}=-\infty$ or setting $\Delta_{upper}=\infty$).}
\label{tab:goertsen}
\end{table}

\section{Conclusion}\label{sec:conclsn}

Researchers require statistical tools that allow them to reject the presence of meaningful effects.  Indeed, such tools are essential to scientific progress; see \citet{serlin1993rational}, \citet{altman1995statistics},  and more recently \citet{amrhein2019scientists}.  {In this paper we considered just such a tool: an equivalence test for linear regression analyses.}  Equivalence tests may improve current research practices by allowing researchers to falsify their predictions concerning the presence of an effect.  In this sense,  equivalence testing provides a more formal approach to the ``good-enough principle'' \citep{serlin1993rational}.

\textcolor{black}{The use of equivalence/non-inferiority tests should not rule out the complementary use of confidence intervals.  Indeed, confidence intervals can be extremely useful for highlighting the stability (or lack of stability) of a given estimator \citep{fidler2004editors}.  One major strength of confidence intervals is that, not only can they indicate if the effect of interest is trivial, but they can also indicate how small the effect may be.  Perhaps one advantage of equivalence/non-inferiority testing over confidence intervals may be that testing can improve the interpretation of null results \citep{parkhurst2001statistical, hauck1986proposal}.  By clearly distinguishing between what is a ``negative'' versus an ``inconclusive'' result, equivalence testing serves to simplify the long ``series of searching questions''  necessary to evaluate a ``failed outcome'' \citep{pocock2016primary}.} {The best interpretation of data might be obtained when using both tools together, or perhaps by reporting the ``least equivalent allowable difference'' (LEAD or ``equivalence confidence interval'') as recommended by \citet{meyners2007least}.}

  Effect sizes need not be dimensionless (or standardized) in order to be meaningful \citep{kelley2012effect}.  However, expanding equivalence testing to standardized effect sizes can help researchers conduct equivalence tests by facilitating what is often a very challenging task: defining an appropriate equivalence margin.  While the use of ``default equivalence margins'' based on standardized effect sizes cannot be whole-heartily recommended for all cases, their use is not unlike the use of ``default priors'' for defining Bayes factors which have indeed proven useful to researchers in many scenarios.  {In the practical examples we showed that testing with Bayes factors and testing with frequentist equivalence tests will often, but not always, lead to similar conclusions. The pros and cons of frequentist versus Bayesian testing methods are a topic of great debate; see \citet{campbell2021re} for an in-depth discussion.}

{Note that our proposed equivalence tests are limited to comparing two models for which the difference in degrees of freedom is 1.  In other words, the tests are not suitable for comparing two nested models where the difference is more than a single variable.  For example, with the salaries data we considered, we cannot use the proposed tests to compare a smaller model with only ``sex'' as a predictor, with a larger model that includes ``sex,'' ``discipline'' and ``rank,'' as predictors.  A more general equivalence test for comparing two nested models will be considered in future work; \citet{tan2012confidence} is an excellent resource for this undertaking.}

%note that we proposed an equivalence test for the semipartial correlation coefficient based in inverting the  adjusted Aloe-Becker large-sample confidence interval.  It would certainly be worthwhile in future research to consider an equivalence test based on alternative approximations for the sampling variability of semipartial correlation coefficients \citep{jones2013computing, yuan2011biases}. 

{We also note that the TOST approach we proposed is not necessarily optimal in the sense that other procedures may have slightly higher power.  For instance, \citet{anderson1983new} proposed the so-called ``power method'' as an alternative to the TOST approach (but note that \citet{frick1987level} and \citet{miiller1990power} expressed concerns that the actual type I error rate of the power method may exceed the nominal level).  More recently, \citet{romano2005optimal} proposed what they call the ``optimal equivalence test'' (based on the folded-Normal distribution) as a more powerful alternative to TOST (see also \citet{mollenhoff2019efficient}).}

Finally, there is certainly potential to expand equivalence
testing for other analyses including for logistic regression and time-to-event models.  {These are objectives for future research and will help to further ``extend the arsenal of confirmatory methods rooted in the frequentist paradigm of inference'' \citep{wellek2017critical}.}

\paragraph*{Available Code -}  All the code used in this paper and relevant materials are made available in an OSF repository: https://osf.io/5yr92/, DOI 10.17605/OSF.IO/5YR92

% We must emphasize that there are many reasons besides the need to overcome arbitrary scales for reporting standardized effects.  For example, \cite{nieminen2013standardised} argue that standardized effect sizes might be helpful for the synthesis of  epidemiological studies.  Standardization can also help with interpretation of a regression analysis: subtracting the mean can improve the interpretation of main effects in the presence of interactions, and dividing by the standard deviation will ensure that all predictors are on a common scale. 
%Let's do the same for linear mixed effects models:https://jonlefcheck.net/2013/03/13/r2-for-linear-mixed-effects-models/

%Equivalence tests may also be useful for balance and placebo tests \cite{hartman2018equivalence}. 

\bibliography{truthinscience} 

\begin{thebibliography}{}

\bibitem [\protect \citeauthoryear {%
Aloe%
\ \BBA {} Becker%
}{%
Aloe%
\ \BBA {} Becker%
}{%
{\protect \APACyear {2012}}%
}]{%
aloe2012effect}
\APACinsertmetastar {%
aloe2012effect}%
\begin{APACrefauthors}%
Aloe, A\BPBI M.%
\BCBT {}\ \BBA {} Becker, B\BPBI J.%
\end{APACrefauthors}%
\unskip\
\newblock
\APACrefYearMonthDay{2012}{}{}.
\newblock
{\BBOQ}\APACrefatitle {An effect size for regression predictors in
  meta-analysis} {An effect size for regression predictors in
  meta-analysis}.{\BBCQ}
\newblock
\APACjournalVolNumPages{Journal of Educational and Behavioral
  Statistics}{37}{2}{278--297}.
\PrintBackRefs{\CurrentBib}

\bibitem [\protect \citeauthoryear {%
Altman%
\ \BBA {} Bland%
}{%
Altman%
\ \BBA {} Bland%
}{%
{\protect \APACyear {1995}}%
}]{%
altman1995statistics}
\APACinsertmetastar {%
altman1995statistics}%
\begin{APACrefauthors}%
Altman, D\BPBI G.%
\BCBT {}\ \BBA {} Bland, J\BPBI M.%
\end{APACrefauthors}%
\unskip\
\newblock
\APACrefYearMonthDay{1995}{}{}.
\newblock
{\BBOQ}\APACrefatitle {Statistics notes: Absence of evidence is not evidence of
  absence} {Statistics notes: Absence of evidence is not evidence of
  absence}.{\BBCQ}
\newblock
\APACjournalVolNumPages{The BMJ}{311}{7003}{485.
  https://doi.org/10.1136/bmj.311.7003.485}.
\PrintBackRefs{\CurrentBib}

\bibitem [\protect \citeauthoryear {%
Amrhein%
, Greenland%
\BCBL {}\ \BBA {} McShane%
}{%
Amrhein%
\ \protect \BOthers {.}}{%
{\protect \APACyear {2019}}%
}]{%
amrhein2019scientists}
\APACinsertmetastar {%
amrhein2019scientists}%
\begin{APACrefauthors}%
Amrhein, V.%
, Greenland, S.%
\BCBL {}\ \BBA {} McShane, B.%
\end{APACrefauthors}%
\unskip\
\newblock
\APACrefYearMonthDay{2019}{}{}.
\newblock
{\BBOQ}\APACrefatitle {Scientists rise up against statistical significance}
  {Scientists rise up against statistical significance}.{\BBCQ}
\newblock
\APACjournalVolNumPages{Nature}{7748}{567}{305--307. doi:
  https://doi.org/10.1038/d41586-019-00857-9}.
\PrintBackRefs{\CurrentBib}

\bibitem [\protect \citeauthoryear {%
Anderson%
\ \BBA {} Hauck%
}{%
Anderson%
\ \BBA {} Hauck%
}{%
{\protect \APACyear {1983}}%
}]{%
anderson1983new}
\APACinsertmetastar {%
anderson1983new}%
\begin{APACrefauthors}%
Anderson, S.%
\BCBT {}\ \BBA {} Hauck, W.%
\end{APACrefauthors}%
\unskip\
\newblock
\APACrefYearMonthDay{1983}{}{}.
\newblock
{\BBOQ}\APACrefatitle {A new procedure for testing equivalence in comparative
  bioavailability and other clinical trials} {A new procedure for testing
  equivalence in comparative bioavailability and other clinical trials}.{\BBCQ}
\newblock
\APACjournalVolNumPages{Communications in {S}tatistics-{T}heory and
  {M}ethods}{12}{23}{2663--2692}.
\PrintBackRefs{\CurrentBib}

\bibitem [\protect \citeauthoryear {%
Azen%
\ \BBA {} Budescu%
}{%
Azen%
\ \BBA {} Budescu%
}{%
{\protect \APACyear {2009}}%
}]{%
azen2009applications}
\APACinsertmetastar {%
azen2009applications}%
\begin{APACrefauthors}%
Azen, R.%
\BCBT {}\ \BBA {} Budescu, D.%
\end{APACrefauthors}%
\unskip\
\newblock
\APACrefYear{2009}.
\newblock
\APACrefbtitle {Applications of multiple regression in psychological research.}
  {Applications of multiple regression in psychological research.}
\newblock
\APACaddressPublisher{}{In R.E. Millsap, A. Maydeu-Olivares (Eds.), The SAGE
  handbook of quantitative methods in psychology (pp. 285--310). Sage Thousand
  Oaks, CA. https://doi.org/10.4135/9780857020994}.
\PrintBackRefs{\CurrentBib}

\bibitem [\protect \citeauthoryear {%
Baguley%
}{%
Baguley%
}{%
{\protect \APACyear {2009}}%
}]{%
baguley2009standardized}
\APACinsertmetastar {%
baguley2009standardized}%
\begin{APACrefauthors}%
Baguley, T.%
\end{APACrefauthors}%
\unskip\
\newblock
\APACrefYearMonthDay{2009}{}{}.
\newblock
{\BBOQ}\APACrefatitle {Standardized or simple effect size: What should be
  reported?} {Standardized or simple effect size: What should be
  reported?}{\BBCQ}
\newblock
\APACjournalVolNumPages{British {J}ournal of {P}sychology}{100}{3}{603--617.
  doi: 10.1348/000712608X377117}.
\PrintBackRefs{\CurrentBib}

\bibitem [\protect \citeauthoryear {%
Batterham%
, Van~Loo%
, Charlton%
, Cliff%
\BCBL {}\ \BBA {} Okely%
}{%
Batterham%
\ \protect \BOthers {.}}{%
{\protect \APACyear {2016}}%
}]{%
batterham2016improved}
\APACinsertmetastar {%
batterham2016improved}%
\begin{APACrefauthors}%
Batterham, M\BPBI J.%
, Van~Loo, C.%
, Charlton, K\BPBI E.%
, Cliff, D\BPBI P.%
\BCBL {}\ \BBA {} Okely, A\BPBI D.%
\end{APACrefauthors}%
\unskip\
\newblock
\APACrefYearMonthDay{2016}{}{}.
\newblock
{\BBOQ}\APACrefatitle {Improved interpretation of studies comparing methods of
  dietary assessment: combining equivalence testing with the limits of
  agreement} {Improved interpretation of studies comparing methods of dietary
  assessment: combining equivalence testing with the limits of
  agreement}.{\BBCQ}
\newblock
\APACjournalVolNumPages{British {J}ournal of {N}utrition}{115}{7}{1273--1280}.
\PrintBackRefs{\CurrentBib}

\bibitem [\protect \citeauthoryear {%
Bedrick%
\ \BBA {} Hund%
}{%
Bedrick%
\ \BBA {} Hund%
}{%
{\protect \APACyear {2018}}%
}]{%
bedrick2018approach}
\APACinsertmetastar {%
bedrick2018approach}%
\begin{APACrefauthors}%
Bedrick, E\BPBI J.%
\BCBT {}\ \BBA {} Hund, L.%
\end{APACrefauthors}%
\unskip\
\newblock
\APACrefYearMonthDay{2018}{}{}.
\newblock
{\BBOQ}\APACrefatitle {An approach for quantifying small effects in regression
  models} {An approach for quantifying small effects in regression
  models}.{\BBCQ}
\newblock
\APACjournalVolNumPages{Statistical {M}ethods in {M}edical
  {R}esearch}{27}{4}{1088--1098}.
\PrintBackRefs{\CurrentBib}

\bibitem [\protect \citeauthoryear {%
Briggs%
, Nguyen%
\BCBL {}\ \BBA {} Trafimow%
}{%
Briggs%
\ \protect \BOthers {.}}{%
{\protect \APACyear {2019}}%
}]{%
briggs2019replacement}
\APACinsertmetastar {%
briggs2019replacement}%
\begin{APACrefauthors}%
Briggs, W\BPBI M.%
, Nguyen, H\BPBI T.%
\BCBL {}\ \BBA {} Trafimow, D.%
\end{APACrefauthors}%
\unskip\
\newblock
\APACrefYearMonthDay{2019}{}{}.
\newblock
{\BBOQ}\APACrefatitle {The replacement for hypothesis testing} {The replacement
  for hypothesis testing}.{\BBCQ}
\newblock
\BIn{} \APACrefbtitle {International {C}onference of the {T}hailand
  {E}conometrics {S}ociety} {International {C}onference of the {T}hailand
  {E}conometrics {S}ociety}\ (\BPGS\ 3--17).
\PrintBackRefs{\CurrentBib}

\bibitem [\protect \citeauthoryear {%
Bring%
}{%
Bring%
}{%
{\protect \APACyear {1994}}%
}]{%
bring1994standardize}
\APACinsertmetastar {%
bring1994standardize}%
\begin{APACrefauthors}%
Bring, J.%
\end{APACrefauthors}%
\unskip\
\newblock
\APACrefYearMonthDay{1994}{}{}.
\newblock
{\BBOQ}\APACrefatitle {How to standardize regression coefficients} {How to
  standardize regression coefficients}.{\BBCQ}
\newblock
\APACjournalVolNumPages{The {A}merican {S}tatistician}{48}{3}{209--213. doi:
  10.1080/00031305.1994.10476059}.
\PrintBackRefs{\CurrentBib}

\bibitem [\protect \citeauthoryear {%
Burgoyne%
, Hambrick%
\BCBL {}\ \BBA {} Macnamara%
}{%
Burgoyne%
\ \protect \BOthers {.}}{%
{\protect \APACyear {2020}}%
}]{%
burgoyne2020firm}
\APACinsertmetastar {%
burgoyne2020firm}%
\begin{APACrefauthors}%
Burgoyne, A\BPBI P.%
, Hambrick, D\BPBI Z.%
\BCBL {}\ \BBA {} Macnamara, B\BPBI N.%
\end{APACrefauthors}%
\unskip\
\newblock
\APACrefYearMonthDay{2020}{}{}.
\newblock
{\BBOQ}\APACrefatitle {How firm are the foundations of mind-set theory? {T}he
  claims appear stronger than the evidence} {How firm are the foundations of
  mind-set theory? {T}he claims appear stronger than the evidence}.{\BBCQ}
\newblock
\APACjournalVolNumPages{Psychological {S}cience}{31}{3}{258--267. doi:
  10.1177/0956797619897588}.
\PrintBackRefs{\CurrentBib}

\bibitem [\protect \citeauthoryear {%
Campbell%
\ \BBA {} Gustafson%
}{%
Campbell%
\ \BBA {} Gustafson%
}{%
{\protect \APACyear {2018}}%
}]{%
campbell2018conditional}
\APACinsertmetastar {%
campbell2018conditional}%
\begin{APACrefauthors}%
Campbell, H.%
\BCBT {}\ \BBA {} Gustafson, P.%
\end{APACrefauthors}%
\unskip\
\newblock
\APACrefYearMonthDay{2018}{}{}.
\newblock
{\BBOQ}\APACrefatitle {Conditional equivalence testing: An alternative remedy
  for publication bias} {Conditional equivalence testing: An alternative remedy
  for publication bias}.{\BBCQ}
\newblock
\APACjournalVolNumPages{{P}{L}o{S} {O}{N}{E}}{13}{4}{e0195145.
  https://doi.org/10.1371/journal.pone.0195145}.
\PrintBackRefs{\CurrentBib}

\bibitem [\protect \citeauthoryear {%
Campbell%
\ \BBA {} Gustafson%
}{%
Campbell%
\ \BBA {} Gustafson%
}{%
{\protect \APACyear {2021}}%
{\protect \APACexlab {{\protect \BCnt {1}}}}}]{%
campbell2021re}
\APACinsertmetastar {%
campbell2021re}%
\begin{APACrefauthors}%
Campbell, H.%
\BCBT {}\ \BBA {} Gustafson, P.%
\end{APACrefauthors}%
\unskip\
\newblock
\APACrefYearMonthDay{2021{\protect \BCnt {1}}}{}{}.
\newblock
{\BBOQ}\APACrefatitle {re: Linde et al.(2021)--The {B}ayes factor,
  {H}{D}{I}-{R}{O}{P}{E} and frequentist equivalence testing are actually all
  equivalent} {re: Linde et al.(2021)--the {B}ayes factor,
  {H}{D}{I}-{R}{O}{P}{E} and frequentist equivalence testing are actually all
  equivalent}.{\BBCQ}
\newblock
\APACjournalVolNumPages{arXiv preprint arXiv:2104.07834 (accepted for
  publication in Psychological Methods)}{}{}{}.
\PrintBackRefs{\CurrentBib}

\bibitem [\protect \citeauthoryear {%
Campbell%
\ \BBA {} Gustafson%
}{%
Campbell%
\ \BBA {} Gustafson%
}{%
{\protect \APACyear {2021}}%
{\protect \APACexlab {{\protect \BCnt {2}}}}}]{%
campbell2018make}
\APACinsertmetastar {%
campbell2018make}%
\begin{APACrefauthors}%
Campbell, H.%
\BCBT {}\ \BBA {} Gustafson, P.%
\end{APACrefauthors}%
\unskip\
\newblock
\APACrefYearMonthDay{2021{\protect \BCnt {2}}}{}{}.
\newblock
{\BBOQ}\APACrefatitle {What to make of equivalence testing with a
  post-specified margin?} {What to make of equivalence testing with a
  post-specified margin?}{\BBCQ}
\newblock
\APACjournalVolNumPages{Meta-{P}sychology}{5. doi:
  https://doi.org/10.15626/MP.2020.2506}{}{}.
\PrintBackRefs{\CurrentBib}

\bibitem [\protect \citeauthoryear {%
Campbell%
\ \BBA {} Gustafson%
}{%
Campbell%
\ \BBA {} Gustafson%
}{%
{\protect \APACyear {2022}}%
}]{%
campbell2022coin}
\APACinsertmetastar {%
campbell2022coin}%
\begin{APACrefauthors}%
Campbell, H.%
\BCBT {}\ \BBA {} Gustafson, P.%
\end{APACrefauthors}%
\unskip\
\newblock
\APACrefYearMonthDay{2022}{}{}.
\newblock
{\BBOQ}\APACrefatitle {Bayes Factors and Posterior Estimation: Two Sides of the
  Very Same Coin} {Bayes factors and posterior estimation: Two sides of the
  very same coin}.{\BBCQ}
\newblock
\APACjournalVolNumPages{The {A}merican {S}tatistician}{0}{0}{1-11}.
\newblock
\begin{APACrefDOI} \doi{10.1080/00031305.2022.2139293} \end{APACrefDOI}
\PrintBackRefs{\CurrentBib}

\bibitem [\protect \citeauthoryear {%
J.~Cohen%
}{%
J.~Cohen%
}{%
{\protect \APACyear {1988}}%
}]{%
cohen1988statistical}
\APACinsertmetastar {%
cohen1988statistical}%
\begin{APACrefauthors}%
Cohen, J.%
\end{APACrefauthors}%
\unskip\
\newblock
\APACrefYear{1988}.
\newblock
\APACrefbtitle {Statistical power analysis for the behavioral sciences}
  {Statistical power analysis for the behavioral sciences}.
\newblock
\APACaddressPublisher{New York, NY}{Routledge {A}cademic}.
\PrintBackRefs{\CurrentBib}

\bibitem [\protect \citeauthoryear {%
P.~Cohen%
, West%
\BCBL {}\ \BBA {} Aiken%
}{%
P.~Cohen%
\ \protect \BOthers {.}}{%
{\protect \APACyear {2013}}%
}]{%
cohen2013applied}
\APACinsertmetastar {%
cohen2013applied}%
\begin{APACrefauthors}%
Cohen, P.%
, West, S\BPBI G.%
\BCBL {}\ \BBA {} Aiken, L\BPBI S.%
\end{APACrefauthors}%
\unskip\
\newblock
\APACrefYear{2013}.
\newblock
\APACrefbtitle {Applied multiple regression/correlation analysis for the
  behavioral sciences} {Applied multiple regression/correlation analysis for
  the behavioral sciences}.
\newblock
\APACaddressPublisher{}{Psychology press}.
\PrintBackRefs{\CurrentBib}

\bibitem [\protect \citeauthoryear {%
Consonni%
\ \BBA {} Veronese%
}{%
Consonni%
\ \BBA {} Veronese%
}{%
{\protect \APACyear {2008}}%
}]{%
consonni2008compatibility}
\APACinsertmetastar {%
consonni2008compatibility}%
\begin{APACrefauthors}%
Consonni, G.%
\BCBT {}\ \BBA {} Veronese, P.%
\end{APACrefauthors}%
\unskip\
\newblock
\APACrefYearMonthDay{2008}{}{}.
\newblock
{\BBOQ}\APACrefatitle {Compatibility of prior specifications across linear
  models} {Compatibility of prior specifications across linear models}.{\BBCQ}
\newblock
\APACjournalVolNumPages{Statistical Science}{23}{3}{332--353. doi:
  10.1214/08-STS258}.
\PrintBackRefs{\CurrentBib}

\bibitem [\protect \citeauthoryear {%
Counsell%
, Cribbie%
\BCBL {}\ \BBA {} Flora%
}{%
Counsell%
\ \protect \BOthers {.}}{%
{\protect \APACyear {2020}}%
}]{%
counsell2020evaluating}
\APACinsertmetastar {%
counsell2020evaluating}%
\begin{APACrefauthors}%
Counsell, A.%
, Cribbie, R\BPBI A.%
\BCBL {}\ \BBA {} Flora, D\BPBI B.%
\end{APACrefauthors}%
\unskip\
\newblock
\APACrefYearMonthDay{2020}{}{}.
\newblock
{\BBOQ}\APACrefatitle {Evaluating equivalence testing methods for measurement
  invariance} {Evaluating equivalence testing methods for measurement
  invariance}.{\BBCQ}
\newblock
\APACjournalVolNumPages{Multivariate {B}ehavioral {R}esearch}{55}{2}{312--328}.
\PrintBackRefs{\CurrentBib}

\bibitem [\protect \citeauthoryear {%
Courville%
\ \BBA {} Thompson%
}{%
Courville%
\ \BBA {} Thompson%
}{%
{\protect \APACyear {2001}}%
}]{%
courville2001use}
\APACinsertmetastar {%
courville2001use}%
\begin{APACrefauthors}%
Courville, T.%
\BCBT {}\ \BBA {} Thompson, B.%
\end{APACrefauthors}%
\unskip\
\newblock
\APACrefYearMonthDay{2001}{}{}.
\newblock
{\BBOQ}\APACrefatitle {Use of structure coefficients in published multiple
  regression articles: $\mathcal{B}$ is not enough} {Use of structure
  coefficients in published multiple regression articles: $\mathcal{B}$ is not
  enough}.{\BBCQ}
\newblock
\APACjournalVolNumPages{Educational and {P}sychological
  {M}easurement}{61}{2}{229--248}.
\PrintBackRefs{\CurrentBib}

\bibitem [\protect \citeauthoryear {%
Darlington%
\ \BBA {} Hayes%
}{%
Darlington%
\ \BBA {} Hayes%
}{%
{\protect \APACyear {2017}}%
}]{%
darlington2017regression}
\APACinsertmetastar {%
darlington2017regression}%
\begin{APACrefauthors}%
Darlington, R\BPBI B.%
\BCBT {}\ \BBA {} Hayes, A\BPBI F.%
\end{APACrefauthors}%
\unskip\
\newblock
\APACrefYearMonthDay{2017}{}{}.
\newblock
{\BBOQ}\APACrefatitle {Regression analysis and linear models} {Regression
  analysis and linear models}.{\BBCQ}
\newblock
\APACjournalVolNumPages{New York, NY: Guilford}{}{}{603--611}.
\PrintBackRefs{\CurrentBib}

\bibitem [\protect \citeauthoryear {%
Disbato%
}{%
Disbato%
}{%
{\protect \APACyear {2016}}%
}]{%
disbato2016}
\APACinsertmetastar {%
disbato2016}%
\begin{APACrefauthors}%
Disbato, D.%
\end{APACrefauthors}%
\unskip\
\newblock
\APACrefYearMonthDay{2016}{}{}.
\newblock
\APACrefbtitle {On effect sizes in multiple regression.} {On effect sizes in
  multiple regression.}
\newblock
\APAChowpublished
  {http://www.daviddisabato.com/blog/2016/4/8/on-effect-sizes-in-multiple-regression}.
\PrintBackRefs{\CurrentBib}

\bibitem [\protect \citeauthoryear {%
Dixon%
\ \BBA {} Pechmann%
}{%
Dixon%
\ \BBA {} Pechmann%
}{%
{\protect \APACyear {2005}}%
}]{%
dixon2005statistical}
\APACinsertmetastar {%
dixon2005statistical}%
\begin{APACrefauthors}%
Dixon, P\BPBI M.%
\BCBT {}\ \BBA {} Pechmann, J\BPBI H.%
\end{APACrefauthors}%
\unskip\
\newblock
\APACrefYearMonthDay{2005}{}{}.
\newblock
{\BBOQ}\APACrefatitle {A statistical test to show negligible trend} {A
  statistical test to show negligible trend}.{\BBCQ}
\newblock
\APACjournalVolNumPages{Ecology}{86}{7}{1751--1756}.
\PrintBackRefs{\CurrentBib}

\bibitem [\protect \citeauthoryear {%
Dixon%
\ \protect \BOthers {.}}{%
Dixon%
\ \protect \BOthers {.}}{%
{\protect \APACyear {2018}}%
}]{%
dixon2018primer}
\APACinsertmetastar {%
dixon2018primer}%
\begin{APACrefauthors}%
Dixon, P\BPBI M.%
, Saint-Maurice, P\BPBI F.%
, Kim, Y.%
, Hibbing, P.%
, Bai, Y.%
\BCBL {}\ \BBA {} Welk, G\BPBI J.%
\end{APACrefauthors}%
\unskip\
\newblock
\APACrefYearMonthDay{2018}{}{}.
\newblock
{\BBOQ}\APACrefatitle {A primer on the use of equivalence testing for
  evaluating measurement agreement} {A primer on the use of equivalence testing
  for evaluating measurement agreement}.{\BBCQ}
\newblock
\APACjournalVolNumPages{Medicine and {S}cience in {S}ports and
  {E}xercise}{50}{4}{837. doi: 10.1249/MSS.0000000000001481}.
\PrintBackRefs{\CurrentBib}

\bibitem [\protect \citeauthoryear {%
Dudgeon%
}{%
Dudgeon%
}{%
{\protect \APACyear {2016}}%
}]{%
dudgeon2016comparative}
\APACinsertmetastar {%
dudgeon2016comparative}%
\begin{APACrefauthors}%
Dudgeon, P.%
\end{APACrefauthors}%
\unskip\
\newblock
\APACrefYearMonthDay{2016}{}{}.
\newblock
{\BBOQ}\APACrefatitle {A comparative investigation of confidence intervals for
  independent variables in linear regression} {A comparative investigation of
  confidence intervals for independent variables in linear regression}.{\BBCQ}
\newblock
\APACjournalVolNumPages{Multivariate {B}ehavioral
  {R}esearch}{51}{2-3}{139--153}.
\PrintBackRefs{\CurrentBib}

\bibitem [\protect \citeauthoryear {%
Etz%
}{%
Etz%
}{%
{\protect \APACyear {2015}}%
}]{%
etz2015using}
\APACinsertmetastar {%
etz2015using}%
\begin{APACrefauthors}%
Etz, A.%
\end{APACrefauthors}%
\unskip\
\newblock
\APACrefYearMonthDay{2015}{}{}.
\newblock
{\BBOQ}\APACrefatitle {Using {B}ayes factors to get the most Out of linear
  regression: {A} practical guide using {R}} {Using {B}ayes factors to get the
  most out of linear regression: {A} practical guide using {R}}.{\BBCQ}
\newblock
\APACjournalVolNumPages{The {W}innower}{}{}{}.
\newblock
\begin{APACrefURL} \url{https://tinyurl.com/2p9a5ymr} \end{APACrefURL}
\PrintBackRefs{\CurrentBib}

\bibitem [\protect \citeauthoryear {%
Fidler%
, Thomason%
, Cumming%
, Finch%
\BCBL {}\ \BBA {} Leeman%
}{%
Fidler%
\ \protect \BOthers {.}}{%
{\protect \APACyear {2004}}%
}]{%
fidler2004editors}
\APACinsertmetastar {%
fidler2004editors}%
\begin{APACrefauthors}%
Fidler, F.%
, Thomason, N.%
, Cumming, G.%
, Finch, S.%
\BCBL {}\ \BBA {} Leeman, J.%
\end{APACrefauthors}%
\unskip\
\newblock
\APACrefYearMonthDay{2004}{}{}.
\newblock
{\BBOQ}\APACrefatitle {Editors can lead researchers to confidence intervals,
  but can't make them think: Statistical reform lessons from medicine} {Editors
  can lead researchers to confidence intervals, but can't make them think:
  Statistical reform lessons from medicine}.{\BBCQ}
\newblock
\APACjournalVolNumPages{Psychological Science}{15}{2}{119--126. doi:
  10.1111/j.0963-7214.2004.01502008.x}.
\PrintBackRefs{\CurrentBib}

\bibitem [\protect \citeauthoryear {%
Fox%
\ \protect \BOthers {.}}{%
Fox%
\ \protect \BOthers {.}}{%
{\protect \APACyear {2012}}%
}]{%
fox2012package}
\APACinsertmetastar {%
fox2012package}%
\begin{APACrefauthors}%
Fox, J.%
, Weisberg, S.%
, Adler, D.%
, Bates, D.%
, Baud-Bovy, G.%
, Ellison, S.%
\BDBL {}Graves, S.%
\end{APACrefauthors}%
\unskip\
\newblock
\APACrefYearMonthDay{2012}{}{}.
\newblock
{\BBOQ}\APACrefatitle {Package ``car''} {Package ``car''}.{\BBCQ}
\newblock
\APACjournalVolNumPages{Vienna: {R} {F}oundation for {S}tatistical
  {C}omputing}{}{}{}.
\PrintBackRefs{\CurrentBib}

\bibitem [\protect \citeauthoryear {%
Fraley%
\ \BBA {} Vazire%
}{%
Fraley%
\ \BBA {} Vazire%
}{%
{\protect \APACyear {2014}}%
}]{%
fraley2014n}
\APACinsertmetastar {%
fraley2014n}%
\begin{APACrefauthors}%
Fraley, R\BPBI C.%
\BCBT {}\ \BBA {} Vazire, S.%
\end{APACrefauthors}%
\unskip\
\newblock
\APACrefYearMonthDay{2014}{}{}.
\newblock
{\BBOQ}\APACrefatitle {The {N}-pact factor: Evaluating the quality of empirical
  journals with respect to sample size and statistical power} {The {N}-pact
  factor: Evaluating the quality of empirical journals with respect to sample
  size and statistical power}.{\BBCQ}
\newblock
\APACjournalVolNumPages{{P}{L}o{S} {O}{N}{E}}{9}{10}{e109019.
  https://doi.org/10.1371/journal.pone.0109019}.
\PrintBackRefs{\CurrentBib}

\bibitem [\protect \citeauthoryear {%
Frick%
}{%
Frick%
}{%
{\protect \APACyear {1987}}%
}]{%
frick1987level}
\APACinsertmetastar {%
frick1987level}%
\begin{APACrefauthors}%
Frick, H.%
\end{APACrefauthors}%
\unskip\
\newblock
\APACrefYearMonthDay{1987}{}{}.
\newblock
{\BBOQ}\APACrefatitle {On level and powee of {A}nderson and {H}auck's peocedure
  for testing equivalence in compative bioavailability} {On level and powee of
  {A}nderson and {H}auck's peocedure for testing equivalence in compative
  bioavailability}.{\BBCQ}
\newblock
\APACjournalVolNumPages{Communications in {S}tatistics-{T}heory and
  {M}ethods}{16}{9}{2771--2778}.
\PrintBackRefs{\CurrentBib}

\bibitem [\protect \citeauthoryear {%
Fruehauf%
, Fair%
, Liebel%
, Bjornn%
\BCBL {}\ \BBA {} Larson%
}{%
Fruehauf%
\ \protect \BOthers {.}}{%
{\protect \APACyear {2021}}%
}]{%
fruehauf2021cognitive}
\APACinsertmetastar {%
fruehauf2021cognitive}%
\begin{APACrefauthors}%
Fruehauf, L\BPBI M.%
, Fair, J\BPBI E.%
, Liebel, S\BPBI W.%
, Bjornn, D.%
\BCBL {}\ \BBA {} Larson, M\BPBI J.%
\end{APACrefauthors}%
\unskip\
\newblock
\APACrefYearMonthDay{2021}{}{}.
\newblock
{\BBOQ}\APACrefatitle {Cognitive control in obsessive-compulsive disorder
  (OCD): Proactive control adjustments or consistent performance?} {Cognitive
  control in obsessive-compulsive disorder (ocd): Proactive control adjustments
  or consistent performance?}{\BBCQ}
\newblock
\APACjournalVolNumPages{Psychiatry Research}{298}{}{113809;
  https://doi.org/10.1016/j.psychres.2021.113809}.
\PrintBackRefs{\CurrentBib}

\bibitem [\protect \citeauthoryear {%
Funder%
\ \BBA {} Ozer%
}{%
Funder%
\ \BBA {} Ozer%
}{%
{\protect \APACyear {2019}}%
}]{%
funder2019evaluating}
\APACinsertmetastar {%
funder2019evaluating}%
\begin{APACrefauthors}%
Funder, D\BPBI C.%
\BCBT {}\ \BBA {} Ozer, D\BPBI J.%
\end{APACrefauthors}%
\unskip\
\newblock
\APACrefYearMonthDay{2019}{}{}.
\newblock
{\BBOQ}\APACrefatitle {Evaluating effect size in psychological research: Sense
  and nonsense} {Evaluating effect size in psychological research: Sense and
  nonsense}.{\BBCQ}
\newblock
\APACjournalVolNumPages{Advances in {M}ethods and {P}ractices in
  {P}sychological {S}cience}{2}{2}{156--168}.
\PrintBackRefs{\CurrentBib}

\bibitem [\protect \citeauthoryear {%
Ghashim%
\ \BBA {} Boily%
}{%
Ghashim%
\ \BBA {} Boily%
}{%
{\protect \APACyear {2018}}%
}]{%
ghashimggplot2}
\APACinsertmetastar {%
ghashimggplot2}%
\begin{APACrefauthors}%
Ghashim, E.%
\BCBT {}\ \BBA {} Boily, P.%
\end{APACrefauthors}%
\unskip\
\newblock
\APACrefYearMonthDay{2018}{}{}.
\newblock
{\BBOQ}\APACrefatitle {A ggplot2 primer} {A ggplot2 primer}.{\BBCQ}
\newblock
\APACjournalVolNumPages{Data Action Lab - {D}ata Science Report Series}{}{}{}.
\newblock
\begin{APACrefURL} \url{https://tinyurl.com/2p9brz9a} \end{APACrefURL}
\PrintBackRefs{\CurrentBib}

\bibitem [\protect \citeauthoryear {%
Goertzen%
\ \BBA {} Cribbie%
}{%
Goertzen%
\ \BBA {} Cribbie%
}{%
{\protect \APACyear {2010}}%
}]{%
goertzen2010detecting}
\APACinsertmetastar {%
goertzen2010detecting}%
\begin{APACrefauthors}%
Goertzen, J\BPBI R.%
\BCBT {}\ \BBA {} Cribbie, R\BPBI A.%
\end{APACrefauthors}%
\unskip\
\newblock
\APACrefYearMonthDay{2010}{}{}.
\newblock
{\BBOQ}\APACrefatitle {Detecting a lack of association: An equivalence testing
  approach} {Detecting a lack of association: An equivalence testing
  approach}.{\BBCQ}
\newblock
\APACjournalVolNumPages{British {J}ournal of {M}athematical and {S}tatistical
  {P}sychology}{63}{3}{527--537. doi: 10.1348/000711009X475853}.
\PrintBackRefs{\CurrentBib}

\bibitem [\protect \citeauthoryear {%
Hartung%
, Cottrell%
\BCBL {}\ \BBA {} Giffin%
}{%
Hartung%
\ \protect \BOthers {.}}{%
{\protect \APACyear {1983}}%
}]{%
hartung1983absence}
\APACinsertmetastar {%
hartung1983absence}%
\begin{APACrefauthors}%
Hartung, J.%
, Cottrell, J\BPBI E.%
\BCBL {}\ \BBA {} Giffin, J\BPBI P.%
\end{APACrefauthors}%
\unskip\
\newblock
\APACrefYearMonthDay{1983}{}{}.
\newblock
{\BBOQ}\APACrefatitle {Absence of evidence is not evidence of absence} {Absence
  of evidence is not evidence of absence}.{\BBCQ}
\newblock
\APACjournalVolNumPages{Anesthesiology: The Journal of the American Society of
  Anesthesiologists}{58}{3}{298--299.
  https://doi.org/10.1097/00000542-198303000-00033}.
\PrintBackRefs{\CurrentBib}

\bibitem [\protect \citeauthoryear {%
Hauck%
\ \BBA {} Anderson%
}{%
Hauck%
\ \BBA {} Anderson%
}{%
{\protect \APACyear {1986}}%
}]{%
hauck1986proposal}
\APACinsertmetastar {%
hauck1986proposal}%
\begin{APACrefauthors}%
Hauck, W\BPBI W.%
\BCBT {}\ \BBA {} Anderson, S.%
\end{APACrefauthors}%
\unskip\
\newblock
\APACrefYearMonthDay{1986}{}{}.
\newblock
{\BBOQ}\APACrefatitle {A proposal for interpreting and reporting negative
  studies} {A proposal for interpreting and reporting negative studies}.{\BBCQ}
\newblock
\APACjournalVolNumPages{Statistics in {M}edicine}{5}{3}{203--209. doi:
  10.1002/sim.4780050302}.
\PrintBackRefs{\CurrentBib}

\bibitem [\protect \citeauthoryear {%
Hemphill%
}{%
Hemphill%
}{%
{\protect \APACyear {2003}}%
}]{%
hemphill2003interpreting}
\APACinsertmetastar {%
hemphill2003interpreting}%
\begin{APACrefauthors}%
Hemphill, J\BPBI F.%
\end{APACrefauthors}%
\unskip\
\newblock
\APACrefYearMonthDay{2003}{}{}.
\newblock
{\BBOQ}\APACrefatitle {Interpreting the Magnitudes of Correlation Coefficients}
  {Interpreting the magnitudes of correlation coefficients}.{\BBCQ}
\newblock
\APACjournalVolNumPages{Journal of Counseling Psychology}{29}{}{58--65.
  https://doi.org/10.1037/0003-066X.58.1.78}.
\PrintBackRefs{\CurrentBib}

\bibitem [\protect \citeauthoryear {%
Hung%
, Wang%
\BCBL {}\ \BBA {} O'Neill%
}{%
Hung%
\ \protect \BOthers {.}}{%
{\protect \APACyear {2005}}%
}]{%
hung2005regulatory}
\APACinsertmetastar {%
hung2005regulatory}%
\begin{APACrefauthors}%
Hung, H.%
, Wang, S\BHBI J.%
\BCBL {}\ \BBA {} O'Neill, R.%
\end{APACrefauthors}%
\unskip\
\newblock
\APACrefYearMonthDay{2005}{}{}.
\newblock
{\BBOQ}\APACrefatitle {A regulatory perspective on choice of margin and
  statistical inference issue in non-inferiority trials} {A regulatory
  perspective on choice of margin and statistical inference issue in
  non-inferiority trials}.{\BBCQ}
\newblock
\APACjournalVolNumPages{Biometrical Journal}{47}{1}{28--36.
  https://doi.org/10.1002/bimj.200410084}.
\PrintBackRefs{\CurrentBib}

\bibitem [\protect \citeauthoryear {%
Kanetkar%
, Evans%
, Everell%
, Irvine%
\BCBL {}\ \BBA {} Millman%
}{%
Kanetkar%
\ \protect \BOthers {.}}{%
{\protect \APACyear {1995}}%
}]{%
kanetkar1995effect}
\APACinsertmetastar {%
kanetkar1995effect}%
\begin{APACrefauthors}%
Kanetkar, V.%
, Evans, M\BPBI G.%
, Everell, S\BPBI A.%
, Irvine, D.%
\BCBL {}\ \BBA {} Millman, Z.%
\end{APACrefauthors}%
\unskip\
\newblock
\APACrefYearMonthDay{1995}{}{}.
\newblock
{\BBOQ}\APACrefatitle {The effect of scale changes on meta-analysis of
  multiplicative and main effects models} {The effect of scale changes on
  meta-analysis of multiplicative and main effects models}.{\BBCQ}
\newblock
\APACjournalVolNumPages{Educational and {P}sychological
  {M}easurement}{55}{2}{206--224}.
\PrintBackRefs{\CurrentBib}

\bibitem [\protect \citeauthoryear {%
Keefe%
\ \protect \BOthers {.}}{%
Keefe%
\ \protect \BOthers {.}}{%
{\protect \APACyear {2013}}%
}]{%
keefe2013defining}
\APACinsertmetastar {%
keefe2013defining}%
\begin{APACrefauthors}%
Keefe, R\BPBI S.%
, Kraemer, H\BPBI C.%
, Epstein, R\BPBI S.%
, Frank, E.%
, Haynes, G.%
, Laughren, T\BPBI P.%
\BDBL {}Leon, A\BPBI C.%
\end{APACrefauthors}%
\unskip\
\newblock
\APACrefYearMonthDay{2013}{}{}.
\newblock
{\BBOQ}\APACrefatitle {Defining a clinically meaningful effect for the design
  and interpretation of randomized controlled trials} {Defining a clinically
  meaningful effect for the design and interpretation of randomized controlled
  trials}.{\BBCQ}
\newblock
\APACjournalVolNumPages{Innovations in Clinical Neuroscience}{10}{5-6 Suppl
  A}{4S--19S. PMCID: PMC3719483}.
\PrintBackRefs{\CurrentBib}

\bibitem [\protect \citeauthoryear {%
Kelley%
\ \BBA {} Preacher%
}{%
Kelley%
\ \BBA {} Preacher%
}{%
{\protect \APACyear {2012}}%
}]{%
kelley2012effect}
\APACinsertmetastar {%
kelley2012effect}%
\begin{APACrefauthors}%
Kelley, K.%
\BCBT {}\ \BBA {} Preacher, K\BPBI J.%
\end{APACrefauthors}%
\unskip\
\newblock
\APACrefYearMonthDay{2012}{}{}.
\newblock
{\BBOQ}\APACrefatitle {On effect size} {On effect size}.{\BBCQ}
\newblock
\APACjournalVolNumPages{Psychological {M}ethods}{17}{2}{137--152.
  https://doi.org/10.1037/a0028086}.
\PrintBackRefs{\CurrentBib}

\bibitem [\protect \citeauthoryear {%
Koh%
\ \BBA {} Cribbie%
}{%
Koh%
\ \BBA {} Cribbie%
}{%
{\protect \APACyear {2013}}%
}]{%
koh2013robust}
\APACinsertmetastar {%
koh2013robust}%
\begin{APACrefauthors}%
Koh, A.%
\BCBT {}\ \BBA {} Cribbie, R.%
\end{APACrefauthors}%
\unskip\
\newblock
\APACrefYearMonthDay{2013}{}{}.
\newblock
{\BBOQ}\APACrefatitle {Robust tests of equivalence for k independent groups}
  {Robust tests of equivalence for k independent groups}.{\BBCQ}
\newblock
\APACjournalVolNumPages{British Journal of Mathematical and Statistical
  Psychology}{66}{3}{426--434. doi: 10.1111/j.2044-8317.2012.02056.x}.
\PrintBackRefs{\CurrentBib}

\bibitem [\protect \citeauthoryear {%
K{\"u}hberger%
, Fritz%
\BCBL {}\ \BBA {} Scherndl%
}{%
K{\"u}hberger%
\ \protect \BOthers {.}}{%
{\protect \APACyear {2014}}%
}]{%
kuhberger2014publication}
\APACinsertmetastar {%
kuhberger2014publication}%
\begin{APACrefauthors}%
K{\"u}hberger, A.%
, Fritz, A.%
\BCBL {}\ \BBA {} Scherndl, T.%
\end{APACrefauthors}%
\unskip\
\newblock
\APACrefYearMonthDay{2014}{}{}.
\newblock
{\BBOQ}\APACrefatitle {Publication bias in psychology: a diagnosis based on the
  correlation between effect size and sample size} {Publication bias in
  psychology: a diagnosis based on the correlation between effect size and
  sample size}.{\BBCQ}
\newblock
\APACjournalVolNumPages{{P}{L}o{S} {O}{N}{E}}{9}{9}{e105825.
  https://doi.org/10.1371/journal.pone.0105825}.
\PrintBackRefs{\CurrentBib}

\bibitem [\protect \citeauthoryear {%
Lakens%
}{%
Lakens%
}{%
{\protect \APACyear {2017}}%
}]{%
lakens2017equivalence}
\APACinsertmetastar {%
lakens2017equivalence}%
\begin{APACrefauthors}%
Lakens, D.%
\end{APACrefauthors}%
\unskip\
\newblock
\APACrefYearMonthDay{2017}{}{}.
\newblock
{\BBOQ}\APACrefatitle {Equivalence tests: a practical primer for t-tests,
  correlations, and meta-analyses} {Equivalence tests: a practical primer for
  t-tests, correlations, and meta-analyses}.{\BBCQ}
\newblock
\APACjournalVolNumPages{Social {P}sychological and {P}ersonality
  {S}cience}{8}{4}{355--362. doi: 10.1177/1948550617697177}.
\PrintBackRefs{\CurrentBib}

\bibitem [\protect \citeauthoryear {%
Lakens%
, Scheel%
\BCBL {}\ \BBA {} Isager%
}{%
Lakens%
\ \protect \BOthers {.}}{%
{\protect \APACyear {2018}}%
}]{%
lakens2018equivalence}
\APACinsertmetastar {%
lakens2018equivalence}%
\begin{APACrefauthors}%
Lakens, D.%
, Scheel, A\BPBI M.%
\BCBL {}\ \BBA {} Isager, P\BPBI M.%
\end{APACrefauthors}%
\unskip\
\newblock
\APACrefYearMonthDay{2018}{}{}.
\newblock
{\BBOQ}\APACrefatitle {Equivalence testing for psychological research: {A}
  tutorial} {Equivalence testing for psychological research: {A}
  tutorial}.{\BBCQ}
\newblock
\APACjournalVolNumPages{Advances in Methods and Practices in Psychological
  Science}{1}{2}{259--269. https://doi.org/10.1177/2515245918770963}.
\PrintBackRefs{\CurrentBib}

\bibitem [\protect \citeauthoryear {%
Leday%
, Engel%
, Vossen%
, de Vos%
\BCBL {}\ \BBA {} van~der Voet%
}{%
Leday%
\ \protect \BOthers {.}}{%
{\protect \APACyear {2022}}%
}]{%
leday2022multivariate}
\APACinsertmetastar {%
leday2022multivariate}%
\begin{APACrefauthors}%
Leday, G\BPBI G.%
, Engel, J.%
, Vossen, J\BPBI H.%
, de Vos, R\BPBI C.%
\BCBL {}\ \BBA {} van~der Voet, H.%
\end{APACrefauthors}%
\unskip\
\newblock
\APACrefYearMonthDay{2022}{}{}.
\newblock
{\BBOQ}\APACrefatitle {Multivariate equivalence testing for food safety
  assessment} {Multivariate equivalence testing for food safety
  assessment}.{\BBCQ}
\newblock
\APACjournalVolNumPages{Food and {C}hemical {T}oxicology}{170}{}{113446}.
\PrintBackRefs{\CurrentBib}

\bibitem [\protect \citeauthoryear {%
Leonidaki%
\ \BBA {} Constantinou%
}{%
Leonidaki%
\ \BBA {} Constantinou%
}{%
{\protect \APACyear {2021}}%
}]{%
leonidaki2021comparison}
\APACinsertmetastar {%
leonidaki2021comparison}%
\begin{APACrefauthors}%
Leonidaki, V.%
\BCBT {}\ \BBA {} Constantinou, M\BPBI P.%
\end{APACrefauthors}%
\unskip\
\newblock
\APACrefYearMonthDay{2021}{}{}.
\newblock
{\BBOQ}\APACrefatitle {A comparison of completion and recovery rates between
  first-line protocol-based cognitive behavioural therapy and non-manualized
  relational therapies within a {U}{K} psychological service} {A comparison of
  completion and recovery rates between first-line protocol-based cognitive
  behavioural therapy and non-manualized relational therapies within a {U}{K}
  psychological service}.{\BBCQ}
\newblock
\APACjournalVolNumPages{Clinical {P}sychology \& {P}sychotherapy}{}{}{1–-13.
  doi: 10.1002/cpp.2669}.
\PrintBackRefs{\CurrentBib}

\bibitem [\protect \citeauthoryear {%
Levine%
, Weber%
, Park%
\BCBL {}\ \BBA {} Hullett%
}{%
Levine%
\ \protect \BOthers {.}}{%
{\protect \APACyear {2008}}%
}]{%
levine2008communication}
\APACinsertmetastar {%
levine2008communication}%
\begin{APACrefauthors}%
Levine, T\BPBI R.%
, Weber, R.%
, Park, H\BPBI S.%
\BCBL {}\ \BBA {} Hullett, C\BPBI R.%
\end{APACrefauthors}%
\unskip\
\newblock
\APACrefYearMonthDay{2008}{}{}.
\newblock
{\BBOQ}\APACrefatitle {A communication researchers’ guide to null hypothesis
  significance testing and alternatives} {A communication researchers’ guide
  to null hypothesis significance testing and alternatives}.{\BBCQ}
\newblock
\APACjournalVolNumPages{Human {C}ommunication {R}esearch}{34}{2}{188--209}.
\PrintBackRefs{\CurrentBib}

\bibitem [\protect \citeauthoryear {%
Liang%
, Paulo%
, Molina%
, Clyde%
\BCBL {}\ \BBA {} Berger%
}{%
Liang%
\ \protect \BOthers {.}}{%
{\protect \APACyear {2008}}%
}]{%
liang2008mixtures}
\APACinsertmetastar {%
liang2008mixtures}%
\begin{APACrefauthors}%
Liang, F.%
, Paulo, R.%
, Molina, G.%
, Clyde, M\BPBI A.%
\BCBL {}\ \BBA {} Berger, J\BPBI O.%
\end{APACrefauthors}%
\unskip\
\newblock
\APACrefYearMonthDay{2008}{}{}.
\newblock
{\BBOQ}\APACrefatitle {Mixtures of $g$-priors for {B}ayesian variable
  selection} {Mixtures of $g$-priors for {B}ayesian variable selection}.{\BBCQ}
\newblock
\APACjournalVolNumPages{Journal of the American Statistical
  Association}{103}{481}{410--423. https://doi.org/10.1198/016214507000001337}.
\PrintBackRefs{\CurrentBib}

\bibitem [\protect \citeauthoryear {%
Lovakov%
\ \BBA {} Agadullina%
}{%
Lovakov%
\ \BBA {} Agadullina%
}{%
{\protect \APACyear {2021}}%
}]{%
lovakov2021empirically}
\APACinsertmetastar {%
lovakov2021empirically}%
\begin{APACrefauthors}%
Lovakov, A.%
\BCBT {}\ \BBA {} Agadullina, E\BPBI R.%
\end{APACrefauthors}%
\unskip\
\newblock
\APACrefYearMonthDay{2021}{}{}.
\newblock
{\BBOQ}\APACrefatitle {Empirically derived guidelines for effect size
  interpretation in social psychology} {Empirically derived guidelines for
  effect size interpretation in social psychology}.{\BBCQ}
\newblock
\APACjournalVolNumPages{European {J}ournal of {S}ocial
  {P}sychology}{51}{3}{485--504}.
\PrintBackRefs{\CurrentBib}

\bibitem [\protect \citeauthoryear {%
Marcoulides%
\ \BBA {} Yuan%
}{%
Marcoulides%
\ \BBA {} Yuan%
}{%
{\protect \APACyear {2017}}%
}]{%
marcoulides2017new}
\APACinsertmetastar {%
marcoulides2017new}%
\begin{APACrefauthors}%
Marcoulides, K\BPBI M.%
\BCBT {}\ \BBA {} Yuan, K\BHBI H.%
\end{APACrefauthors}%
\unskip\
\newblock
\APACrefYearMonthDay{2017}{}{}.
\newblock
{\BBOQ}\APACrefatitle {New ways to evaluate goodness of fit: A note on using
  equivalence testing to assess structural equation models} {New ways to
  evaluate goodness of fit: A note on using equivalence testing to assess
  structural equation models}.{\BBCQ}
\newblock
\APACjournalVolNumPages{Structural Equation Modeling: A Multidisciplinary
  Journal}{24}{1}{148--153}.
\PrintBackRefs{\CurrentBib}

\bibitem [\protect \citeauthoryear {%
Marszalek%
, Barber%
, Kohlhart%
\BCBL {}\ \BBA {} Cooper%
}{%
Marszalek%
\ \protect \BOthers {.}}{%
{\protect \APACyear {2011}}%
}]{%
marszalek2011sample}
\APACinsertmetastar {%
marszalek2011sample}%
\begin{APACrefauthors}%
Marszalek, J\BPBI M.%
, Barber, C.%
, Kohlhart, J.%
\BCBL {}\ \BBA {} Cooper, B\BPBI H.%
\end{APACrefauthors}%
\unskip\
\newblock
\APACrefYearMonthDay{2011}{}{}.
\newblock
{\BBOQ}\APACrefatitle {Sample size in psychological research over the past 30
  years} {Sample size in psychological research over the past 30 years}.{\BBCQ}
\newblock
\APACjournalVolNumPages{Perceptual and Motor Skills}{112}{2}{331--348. doi:
  10.2466/03.11.PMS.112.2.331-348}.
\PrintBackRefs{\CurrentBib}

\bibitem [\protect \citeauthoryear {%
Meyners%
}{%
Meyners%
}{%
{\protect \APACyear {2007}}%
}]{%
meyners2007least}
\APACinsertmetastar {%
meyners2007least}%
\begin{APACrefauthors}%
Meyners, M.%
\end{APACrefauthors}%
\unskip\
\newblock
\APACrefYearMonthDay{2007}{}{}.
\newblock
{\BBOQ}\APACrefatitle {Least equivalent allowable differences in equivalence
  testing} {Least equivalent allowable differences in equivalence
  testing}.{\BBCQ}
\newblock
\APACjournalVolNumPages{Food Quality and Preference}{18}{3}{541--547}.
\PrintBackRefs{\CurrentBib}

\bibitem [\protect \citeauthoryear {%
M{\"o}llenhoff%
\ \protect \BOthers {.}}{%
M{\"o}llenhoff%
\ \protect \BOthers {.}}{%
{\protect \APACyear {2022}}%
}]{%
mollenhoff2019efficient}
\APACinsertmetastar {%
mollenhoff2019efficient}%
\begin{APACrefauthors}%
M{\"o}llenhoff, K.%
, Loingeville, F.%
, Bertrand, J.%
, Nguyen, T\BPBI T.%
, Sharan, S.%
, Zhao, L.%
\BDBL {}Mentr{\'e}, F.%
\end{APACrefauthors}%
\unskip\
\newblock
\APACrefYearMonthDay{2022}{}{}.
\newblock
{\BBOQ}\APACrefatitle {Efficient model-based bioequivalence testing} {Efficient
  model-based bioequivalence testing}.{\BBCQ}
\newblock
\APACjournalVolNumPages{Biostatistics}{23}{1}{314--327. doi:
  10.1093/biostatistics/kxaa026}.
\PrintBackRefs{\CurrentBib}

\bibitem [\protect \citeauthoryear {%
Moon%
}{%
Moon%
}{%
{\protect \APACyear {2017}}%
}]{%
moon2017learn}
\APACinsertmetastar {%
moon2017learn}%
\begin{APACrefauthors}%
Moon, K\BHBI W.%
\end{APACrefauthors}%
\unskip\
\newblock
\APACrefYear{2017}.
\newblock
\APACrefbtitle {Learn ``ggplot2'' {U}sing {S}hiny {A}pp} {Learn ``ggplot2''
  {U}sing {S}hiny {A}pp}.
\newblock
\APACaddressPublisher{Cham, Switzerland}{Springer {I}nternational {P}ublishing.
  doi: 10.1007/978-3-319-53019-2}.
\PrintBackRefs{\CurrentBib}

\bibitem [\protect \citeauthoryear {%
Morey%
\ \BBA {} Rouder%
}{%
Morey%
\ \BBA {} Rouder%
}{%
{\protect \APACyear {2011}}%
}]{%
morey2011bayes}
\APACinsertmetastar {%
morey2011bayes}%
\begin{APACrefauthors}%
Morey, R\BPBI D.%
\BCBT {}\ \BBA {} Rouder, J\BPBI N.%
\end{APACrefauthors}%
\unskip\
\newblock
\APACrefYearMonthDay{2011}{}{}.
\newblock
{\BBOQ}\APACrefatitle {{B}ayes factor approaches for testing interval null
  hypotheses.} {{B}ayes factor approaches for testing interval null
  hypotheses.}{\BBCQ}
\newblock
\APACjournalVolNumPages{Psychological Methods}{16}{4}{406--419. doi:
  10.1037/a0024377}.
\PrintBackRefs{\CurrentBib}

\bibitem [\protect \citeauthoryear {%
Morey%
, Rouder%
, Jamil%
\BCBL {}\ \BBA {} Morey%
}{%
Morey%
\ \protect \BOthers {.}}{%
{\protect \APACyear {2015}}%
}]{%
morey2015package}
\APACinsertmetastar {%
morey2015package}%
\begin{APACrefauthors}%
Morey, R\BPBI D.%
, Rouder, J\BPBI N.%
, Jamil, T.%
\BCBL {}\ \BBA {} Morey, M\BPBI R\BPBI D.%
\end{APACrefauthors}%
\unskip\
\newblock
\APACrefYearMonthDay{2015}{}{}.
\newblock
{\BBOQ}\APACrefatitle {Package `{B}ayes{F}actor'} {Package
  `{B}ayes{F}actor'}.{\BBCQ}
\newblock
\APACjournalVolNumPages{CRAN repository}{}{}{}.
\newblock
\begin{APACrefURL}
  \url{https://cran.r-project.org/web/packages/BayesFactor/index.html}
  \end{APACrefURL}
\PrintBackRefs{\CurrentBib}

\bibitem [\protect \citeauthoryear {%
Morris%
, White%
\BCBL {}\ \BBA {} Crowther%
}{%
Morris%
\ \protect \BOthers {.}}{%
{\protect \APACyear {2019}}%
}]{%
morris2019using}
\APACinsertmetastar {%
morris2019using}%
\begin{APACrefauthors}%
Morris, T\BPBI P.%
, White, I\BPBI R.%
\BCBL {}\ \BBA {} Crowther, M\BPBI J.%
\end{APACrefauthors}%
\unskip\
\newblock
\APACrefYearMonthDay{2019}{}{}.
\newblock
{\BBOQ}\APACrefatitle {Using simulation studies to evaluate statistical
  methods} {Using simulation studies to evaluate statistical methods}.{\BBCQ}
\newblock
\APACjournalVolNumPages{Statistics in {M}edicine}{38}{11}{2074--2102.
  https://doi.org/10.1002/sim.8086}.
\PrintBackRefs{\CurrentBib}

\bibitem [\protect \citeauthoryear {%
M{\"u}ller-Coors%
}{%
M{\"u}ller-Coors%
}{%
{\protect \APACyear {1990}}%
}]{%
miiller1990power}
\APACinsertmetastar {%
miiller1990power}%
\begin{APACrefauthors}%
M{\"u}ller-Coors, J.%
\end{APACrefauthors}%
\unskip\
\newblock
\APACrefYearMonthDay{1990}{}{}.
\newblock
{\BBOQ}\APACrefatitle {The power of the {A}nderson-{H}auck test and the
  double-test} {The power of the {A}nderson-{H}auck test and the
  double-test}.{\BBCQ}
\newblock
\APACjournalVolNumPages{Biometrical {J}ournal}{32}{}{259--266}.
\PrintBackRefs{\CurrentBib}

\bibitem [\protect \citeauthoryear {%
Parkhurst%
}{%
Parkhurst%
}{%
{\protect \APACyear {2001}}%
}]{%
parkhurst2001statistical}
\APACinsertmetastar {%
parkhurst2001statistical}%
\begin{APACrefauthors}%
Parkhurst, D\BPBI F.%
\end{APACrefauthors}%
\unskip\
\newblock
\APACrefYearMonthDay{2001}{}{}.
\newblock
{\BBOQ}\APACrefatitle {Statistical Significance Tests: Equivalence and Reverse
  Tests Should Reduce Misinterpretation} {Statistical significance tests:
  Equivalence and reverse tests should reduce misinterpretation}.{\BBCQ}
\newblock
\APACjournalVolNumPages{Bioscience}{51}{12}{1051--1057.
  https://doi.org/10.1641/0006-3568(2001)051[1051:SSTEAR]2.0.CO;2}.
\PrintBackRefs{\CurrentBib}

\bibitem [\protect \citeauthoryear {%
Pocock%
\ \BBA {} Stone%
}{%
Pocock%
\ \BBA {} Stone%
}{%
{\protect \APACyear {2016}}%
}]{%
pocock2016primary}
\APACinsertmetastar {%
pocock2016primary}%
\begin{APACrefauthors}%
Pocock, S\BPBI J.%
\BCBT {}\ \BBA {} Stone, G\BPBI W.%
\end{APACrefauthors}%
\unskip\
\newblock
\APACrefYearMonthDay{2016}{}{}.
\newblock
{\BBOQ}\APACrefatitle {The primary outcome fails -what next?} {The primary
  outcome fails -what next?}{\BBCQ}
\newblock
\APACjournalVolNumPages{New England Journal of Medicine}{375}{9}{861--870. doi:
  10.1016/j.jacc.2021.06.024}.
\PrintBackRefs{\CurrentBib}

\bibitem [\protect \citeauthoryear {%
{R Core Team}%
}{%
{R Core Team}%
}{%
{\protect \APACyear {2020}}%
}]{%
rsoftware}
\APACinsertmetastar {%
rsoftware}%
\begin{APACrefauthors}%
{R Core Team}.%
\end{APACrefauthors}%
\unskip\
\newblock
\APACrefYearMonthDay{2020}{}{}.
\newblock
{\BBOQ}\APACrefatitle {R: A Language and Environment for Statistical Computing}
  {R: A language and environment for statistical computing}{\BBCQ}\
  [\bibcomputersoftwaremanual].
\newblock
\APACaddressPublisher{Vienna, Austria}{}.
\newblock
\begin{APACrefURL} \url{\url{https://www.R-project.org/}} \end{APACrefURL}
\PrintBackRefs{\CurrentBib}

\bibitem [\protect \citeauthoryear {%
Richard%
, Bond~Jr%
\BCBL {}\ \BBA {} Stokes-Zoota%
}{%
Richard%
\ \protect \BOthers {.}}{%
{\protect \APACyear {2003}}%
}]{%
richard2003one}
\APACinsertmetastar {%
richard2003one}%
\begin{APACrefauthors}%
Richard, F\BPBI D.%
, Bond~Jr, C\BPBI F.%
\BCBL {}\ \BBA {} Stokes-Zoota, J\BPBI J.%
\end{APACrefauthors}%
\unskip\
\newblock
\APACrefYearMonthDay{2003}{}{}.
\newblock
\APACrefbtitle {One hundred years of social psychology quantitatively
  described.} {One hundred years of social psychology quantitatively
  described.}
\newblock
\APACaddressPublisher{}{Educational Publishing Foundation}.
\PrintBackRefs{\CurrentBib}

\bibitem [\protect \citeauthoryear {%
Robinson%
\ \BBA {} Robinson%
}{%
Robinson%
\ \BBA {} Robinson%
}{%
{\protect \APACyear {2016}}%
}]{%
robinson2016package}
\APACinsertmetastar {%
robinson2016package}%
\begin{APACrefauthors}%
Robinson, A.%
\BCBT {}\ \BBA {} Robinson, M\BPBI A.%
\end{APACrefauthors}%
\unskip\
\newblock
\APACrefYearMonthDay{2016}{}{}.
\newblock
\APACrefbtitle {C{R}{A}{N} Repository: Package `equivalence'.} {C{R}{A}{N}
  repository: Package `equivalence'.}
\newblock
\APAChowpublished {\url{https://tinyurl.com/34brwssa}}.
\PrintBackRefs{\CurrentBib}

\bibitem [\protect \citeauthoryear {%
Rogers%
, Howard%
\BCBL {}\ \BBA {} Vessey%
}{%
Rogers%
\ \protect \BOthers {.}}{%
{\protect \APACyear {1993}}%
}]{%
rogers1993using}
\APACinsertmetastar {%
rogers1993using}%
\begin{APACrefauthors}%
Rogers, J\BPBI L.%
, Howard, K\BPBI I.%
\BCBL {}\ \BBA {} Vessey, J\BPBI T.%
\end{APACrefauthors}%
\unskip\
\newblock
\APACrefYearMonthDay{1993}{}{}.
\newblock
{\BBOQ}\APACrefatitle {Using significance tests to evaluate equivalence between
  two experimental groups.} {Using significance tests to evaluate equivalence
  between two experimental groups.}{\BBCQ}
\newblock
\APACjournalVolNumPages{Psychological {B}ulletin}{113}{3}{553}.
\PrintBackRefs{\CurrentBib}

\bibitem [\protect \citeauthoryear {%
Romano%
}{%
Romano%
}{%
{\protect \APACyear {2005}}%
}]{%
romano2005optimal}
\APACinsertmetastar {%
romano2005optimal}%
\begin{APACrefauthors}%
Romano, J\BPBI P.%
\end{APACrefauthors}%
\unskip\
\newblock
\APACrefYearMonthDay{2005}{}{}.
\newblock
{\BBOQ}\APACrefatitle {Optimal testing of equivalence hypotheses} {Optimal
  testing of equivalence hypotheses}.{\BBCQ}
\newblock
\APACjournalVolNumPages{The {A}nnals of {S}tatistics}{33}{3}{1036--1047}.
\PrintBackRefs{\CurrentBib}

\bibitem [\protect \citeauthoryear {%
Rouder%
\ \BBA {} Morey%
}{%
Rouder%
\ \BBA {} Morey%
}{%
{\protect \APACyear {2012}}%
}]{%
rouder2012default}
\APACinsertmetastar {%
rouder2012default}%
\begin{APACrefauthors}%
Rouder, J\BPBI N.%
\BCBT {}\ \BBA {} Morey, R\BPBI D.%
\end{APACrefauthors}%
\unskip\
\newblock
\APACrefYearMonthDay{2012}{}{}.
\newblock
{\BBOQ}\APACrefatitle {Default {B}ayes factors for model selection in
  regression} {Default {B}ayes factors for model selection in
  regression}.{\BBCQ}
\newblock
\APACjournalVolNumPages{Multivariate Behavioral Research}{47}{6}{877--903.
  https://doi.org/10.1080/00273171.2012.734737}.
\PrintBackRefs{\CurrentBib}

\bibitem [\protect \citeauthoryear {%
Schuirmann%
}{%
Schuirmann%
}{%
{\protect \APACyear {1987}}%
}]{%
schuirmann1987comparison}
\APACinsertmetastar {%
schuirmann1987comparison}%
\begin{APACrefauthors}%
Schuirmann, D\BPBI J.%
\end{APACrefauthors}%
\unskip\
\newblock
\APACrefYearMonthDay{1987}{}{}.
\newblock
{\BBOQ}\APACrefatitle {A comparison of the two one-sided tests procedure and
  the power approach for assessing the equivalence of average bioavailability}
  {A comparison of the two one-sided tests procedure and the power approach for
  assessing the equivalence of average bioavailability}.{\BBCQ}
\newblock
\APACjournalVolNumPages{Journal of pharmacokinetics and
  biopharmaceutics}{15}{6}{657--680}.
\PrintBackRefs{\CurrentBib}

\bibitem [\protect \citeauthoryear {%
Seaman%
\ \BBA {} Serlin%
}{%
Seaman%
\ \BBA {} Serlin%
}{%
{\protect \APACyear {1998}}%
}]{%
seaman1998equivalence}
\APACinsertmetastar {%
seaman1998equivalence}%
\begin{APACrefauthors}%
Seaman, M\BPBI A.%
\BCBT {}\ \BBA {} Serlin, R\BPBI C.%
\end{APACrefauthors}%
\unskip\
\newblock
\APACrefYearMonthDay{1998}{}{}.
\newblock
{\BBOQ}\APACrefatitle {Equivalence confidence intervals for two-group
  comparisons of means.} {Equivalence confidence intervals for two-group
  comparisons of means.}{\BBCQ}
\newblock
\APACjournalVolNumPages{Psychological {M}ethods}{3}{4}{403--411.
  https://doi.org/10.1037/1082-989X.3.4.403}.
\PrintBackRefs{\CurrentBib}

\bibitem [\protect \citeauthoryear {%
Serlin%
, Lapsley%
, Keren%
\BCBL {}\ \BBA {} Lewis%
}{%
Serlin%
\ \protect \BOthers {.}}{%
{\protect \APACyear {1993}}%
}]{%
serlin1993rational}
\APACinsertmetastar {%
serlin1993rational}%
\begin{APACrefauthors}%
Serlin, R\BPBI C.%
, Lapsley, D\BPBI K.%
, Keren, G.%
\BCBL {}\ \BBA {} Lewis, C.%
\end{APACrefauthors}%
\unskip\
\newblock
\APACrefYearMonthDay{1993}{}{}.
\newblock
{\BBOQ}\APACrefatitle {Rational appraisal of psychological research and the
  good-enough principle} {Rational appraisal of psychological research and the
  good-enough principle}.{\BBCQ}
\newblock
\BIn{} G.~Keren\ \BBA {} C.~Lewis\ (\BEDS), \APACrefbtitle {{A} handbook for
  data analysis in the behavioral sciences: {M}ethodological issues} {{A}
  handbook for data analysis in the behavioral sciences: {M}ethodological
  issues}\ (\BPGS\ 199--228).
\newblock
\APACaddressPublisher{Hillsdale, NJ}{Psychology {P}ress}.
\PrintBackRefs{\CurrentBib}

\bibitem [\protect \citeauthoryear {%
Shen%
}{%
Shen%
}{%
{\protect \APACyear {2023}}%
}]{%
shen2023equivalence}
\APACinsertmetastar {%
shen2023equivalence}%
\begin{APACrefauthors}%
Shen, P\BHBI S.%
\end{APACrefauthors}%
\unskip\
\newblock
\APACrefYearMonthDay{2023}{}{}.
\newblock
{\BBOQ}\APACrefatitle {Equivalence tests before end of follow-up under the
  class of log transformation model} {Equivalence tests before end of follow-up
  under the class of log transformation model}.{\BBCQ}
\newblock
\APACjournalVolNumPages{Journal of {B}iopharmaceutical
  Statistics}{33}{3}{324--334}.
\PrintBackRefs{\CurrentBib}

\bibitem [\protect \citeauthoryear {%
Tan~Jr%
}{%
Tan~Jr%
}{%
{\protect \APACyear {2012}}%
}]{%
tan2012confidence}
\APACinsertmetastar {%
tan2012confidence}%
\begin{APACrefauthors}%
Tan~Jr, L.%
\end{APACrefauthors}%
\unskip\
\newblock
\APACrefYearMonthDay{2012}{}{}.
\newblock
{\BBOQ}\APACrefatitle {Confidence Intervals for Comparison of the Squared
  Multiple Correlation Coefficients of Non-nested Models} {Confidence intervals
  for comparison of the squared multiple correlation coefficients of non-nested
  models}.{\BBCQ}
\newblock
\APACjournalVolNumPages{Thesis submitted in partial fulfillment of the
  requirements for the degree in Master of Science; {T}he {U}niversity of
  {W}estern {O}ntario}{}{}{}.
\PrintBackRefs{\CurrentBib}

\bibitem [\protect \citeauthoryear {%
Tendeiro%
\ \BBA {} Kiers%
}{%
Tendeiro%
\ \BBA {} Kiers%
}{%
{\protect \APACyear {2019}}%
}]{%
tendeiro2019review}
\APACinsertmetastar {%
tendeiro2019review}%
\begin{APACrefauthors}%
Tendeiro, J\BPBI N.%
\BCBT {}\ \BBA {} Kiers, H\BPBI A.%
\end{APACrefauthors}%
\unskip\
\newblock
\APACrefYearMonthDay{2019}{}{}.
\newblock
{\BBOQ}\APACrefatitle {A review of issues about null hypothesis {B}ayesian
  testing} {A review of issues about null hypothesis {B}ayesian
  testing}.{\BBCQ}
\newblock
\APACjournalVolNumPages{Psychological {M}ethods}{24}{6}{774--795. doi:
  10.1037/met0000221}.
\PrintBackRefs{\CurrentBib}

\bibitem [\protect \citeauthoryear {%
Tonidandel%
\ \BBA {} LeBreton%
}{%
Tonidandel%
\ \BBA {} LeBreton%
}{%
{\protect \APACyear {2011}}%
}]{%
tonidandel2011relative}
\APACinsertmetastar {%
tonidandel2011relative}%
\begin{APACrefauthors}%
Tonidandel, S.%
\BCBT {}\ \BBA {} LeBreton, J\BPBI M.%
\end{APACrefauthors}%
\unskip\
\newblock
\APACrefYearMonthDay{2011}{}{}.
\newblock
{\BBOQ}\APACrefatitle {Relative importance analysis: A useful supplement to
  regression analysis} {Relative importance analysis: A useful supplement to
  regression analysis}.{\BBCQ}
\newblock
\APACjournalVolNumPages{Journal of {B}usiness and {P}sychology}{26}{1}{1--9}.
\PrintBackRefs{\CurrentBib}

\bibitem [\protect \citeauthoryear {%
Weber%
\ \BBA {} Popova%
}{%
Weber%
\ \BBA {} Popova%
}{%
{\protect \APACyear {2012}}%
}]{%
weber2012testing}
\APACinsertmetastar {%
weber2012testing}%
\begin{APACrefauthors}%
Weber, R.%
\BCBT {}\ \BBA {} Popova, L.%
\end{APACrefauthors}%
\unskip\
\newblock
\APACrefYearMonthDay{2012}{}{}.
\newblock
{\BBOQ}\APACrefatitle {Testing equivalence in communication research: Theory
  and application} {Testing equivalence in communication research: Theory and
  application}.{\BBCQ}
\newblock
\APACjournalVolNumPages{Communication {M}ethods and {M}easures}{6}{3}{190--213.
  https://doi.org/10.1080/19312458.2012.703834}.
\PrintBackRefs{\CurrentBib}

\bibitem [\protect \citeauthoryear {%
Wellek%
}{%
Wellek%
}{%
{\protect \APACyear {2010}}%
}]{%
wellek2010testing}
\APACinsertmetastar {%
wellek2010testing}%
\begin{APACrefauthors}%
Wellek, S.%
\end{APACrefauthors}%
\unskip\
\newblock
\APACrefYear{2010}.
\newblock
\APACrefbtitle {Testing statistical hypotheses of equivalence and
  noninferiority} {Testing statistical hypotheses of equivalence and
  noninferiority}.
\newblock
\APACaddressPublisher{Boca Raton, FL}{Chapman and Hall/CRC}.
\PrintBackRefs{\CurrentBib}

\bibitem [\protect \citeauthoryear {%
Wellek%
}{%
Wellek%
}{%
{\protect \APACyear {2017}}%
}]{%
wellek2017critical}
\APACinsertmetastar {%
wellek2017critical}%
\begin{APACrefauthors}%
Wellek, S.%
\end{APACrefauthors}%
\unskip\
\newblock
\APACrefYearMonthDay{2017}{}{}.
\newblock
{\BBOQ}\APACrefatitle {A critical evaluation of the current ``p-value
  controversy''} {A critical evaluation of the current ``p-value
  controversy''}.{\BBCQ}
\newblock
\APACjournalVolNumPages{Biometrical {J}ournal}{59}{5}{854--872. doi:
  10.1002/bimj.201700001}.
\PrintBackRefs{\CurrentBib}

\bibitem [\protect \citeauthoryear {%
West%
, Aiken%
, Wu%
\BCBL {}\ \BBA {} Taylor%
}{%
West%
\ \protect \BOthers {.}}{%
{\protect \APACyear {2007}}%
}]{%
west2007multiple}
\APACinsertmetastar {%
west2007multiple}%
\begin{APACrefauthors}%
West, S\BPBI G.%
, Aiken, L\BPBI S.%
, Wu, W.%
\BCBL {}\ \BBA {} Taylor, A\BPBI B.%
\end{APACrefauthors}%
\unskip\
\newblock
\APACrefYearMonthDay{2007}{}{}.
\newblock
{\BBOQ}\APACrefatitle {Multiple regression: {A}pplications of the basics and
  beyond in personality research} {Multiple regression: {A}pplications of the
  basics and beyond in personality research}.{\BBCQ}
\newblock
\BIn{} R\BPBI W.~Robins, R\BPBI C.~Fraley\BCBL {}\ \BBA {} R\BPBI F.~Krueger\
  (\BEDS), \APACrefbtitle {Handbook of research methods in personality
  psychology} {Handbook of research methods in personality psychology}\ (\BPGS\
  573--601).
\newblock
\APACaddressPublisher{}{The {G}uilford {P}ress}.
\PrintBackRefs{\CurrentBib}

\bibitem [\protect \citeauthoryear {%
Westlake%
}{%
Westlake%
}{%
{\protect \APACyear {1972}}%
}]{%
westlake1972use}
\APACinsertmetastar {%
westlake1972use}%
\begin{APACrefauthors}%
Westlake, W\BPBI J.%
\end{APACrefauthors}%
\unskip\
\newblock
\APACrefYearMonthDay{1972}{}{}.
\newblock
{\BBOQ}\APACrefatitle {Use of confidence intervals in analysis of comparative
  bioavailability trials} {Use of confidence intervals in analysis of
  comparative bioavailability trials}.{\BBCQ}
\newblock
\APACjournalVolNumPages{Journal of Pharmaceutical Sciences}{61}{8}{1340--1341}.
\PrintBackRefs{\CurrentBib}

\bibitem [\protect \citeauthoryear {%
Wiens%
}{%
Wiens%
}{%
{\protect \APACyear {2002}}%
}]{%
wiens2002choosing}
\APACinsertmetastar {%
wiens2002choosing}%
\begin{APACrefauthors}%
Wiens, B\BPBI L.%
\end{APACrefauthors}%
\unskip\
\newblock
\APACrefYearMonthDay{2002}{}{}.
\newblock
{\BBOQ}\APACrefatitle {Choosing an equivalence limit for noninferiority or
  equivalence studies} {Choosing an equivalence limit for noninferiority or
  equivalence studies}.{\BBCQ}
\newblock
\APACjournalVolNumPages{Controlled Clinical Trials}{23}{1}{2--14. doi:
  10.1016/s0197-2456(01)00196-9}.
\PrintBackRefs{\CurrentBib}

\bibitem [\protect \citeauthoryear {%
Wilkinson%
}{%
Wilkinson%
}{%
{\protect \APACyear {1999}}%
}]{%
wilkinson1999statistical}
\APACinsertmetastar {%
wilkinson1999statistical}%
\begin{APACrefauthors}%
Wilkinson, L.%
\end{APACrefauthors}%
\unskip\
\newblock
\APACrefYearMonthDay{1999}{}{}.
\newblock
{\BBOQ}\APACrefatitle {Statistical methods in psychology journals: {G}uidelines
  and explanations.} {Statistical methods in psychology journals: {G}uidelines
  and explanations.}{\BBCQ}
\newblock
\APACjournalVolNumPages{American {P}sychologist}{54}{8}{594}.
\PrintBackRefs{\CurrentBib}

\bibitem [\protect \citeauthoryear {%
Zhang%
}{%
Zhang%
}{%
{\protect \APACyear {2003}}%
}]{%
zhang2003simple}
\APACinsertmetastar {%
zhang2003simple}%
\begin{APACrefauthors}%
Zhang, P.%
\end{APACrefauthors}%
\unskip\
\newblock
\APACrefYearMonthDay{2003}{}{}.
\newblock
{\BBOQ}\APACrefatitle {A simple formula for sample size calculation in
  equivalence studies} {A simple formula for sample size calculation in
  equivalence studies}.{\BBCQ}
\newblock
\APACjournalVolNumPages{Journal of {B}iopharmaceutical
  {S}tatistics}{13}{3}{529--538}.
\PrintBackRefs{\CurrentBib}

\end{thebibliography}

\pagebreak
		
 \section*{{Supplemental Material}}

 \subsection*{Additional formulas, notation and tables}
 
 \subsubsection*{Least squares estimation}
 
 For completeness, we provide details and notation for least squares estimation in a standard linear regression model.  We define:
\begin{equation}
\reallywidehat{\beta}_{k} = ((X^{T}X)^{-1}X^{T}y)_{k}, \textrm{for $k$ in 1,..., $K$, and}
\label{eq:beta}
\end{equation}
\begin{equation}  \hat{\sigma} = \sqrt{\sum_{i=1}^{N}(\hat{\epsilon}_{i}^2)/(N-K-1)},
\label{eq:sigma}
\end{equation}
\noindent where $\hat{\epsilon}_{i}= \reallywidehat{y}_{i} -{y}_{i}$, and $\reallywidehat{y}_{i} = X_{i \times}^{T}\reallywidehat{\beta}$, for $i$ in 1,..., $N$. { We also define $R^{2}_{YX}$, the coefficient of determination from the linear regression of $Y$ predicted from $X$:}
\begin{equation} \label{eq:P2}
R^{2}_{YX} = \frac{\sigma_{XY}^{T}\Sigma^{-1}_{X}\sigma_{XY}}{\sigma^{2}_{Y}},
\end{equation}
\noindent {where $\sigma^{2}_{Y}=(\beta^{T}\textrm{Cov}(X)\beta + \sigma^{2})$ is the unconditional variance of $Y$, (note that: $\sigma^{2}_{Y} \ge \sigma^{2}$);  $\sigma_{XY}$ is the vector of population covariances between the $K$ different predictors and $Y$; and $\Sigma_{X}$ is the population covariance matrix of the $K$ different predictors.  The $\hat{R}^{2}_{YX}$ statistic estimates the parameter $R^{2}_{YX}$ from the observed data:}

\begin{equation} 
 \hat{R}^{2}_{YX} = 1- \frac{{\sum_{i=1}^{N}\hat{\epsilon}_{i}^2}}{{\sum_{i=1}^{N}({y}_{i}-\bar{y})^2}}, 
 \label{eq:rsquared}
\end{equation}
\noindent {where $\bar{y}= \sum_{i=1}^{N}{y_{i}}/N$.}

A standard NHST for the $k$-th predictor, $X_{k}$, is stated as:

\noindent $\textrm{H}_{0}:  \beta_{k} = 0 $, vs. \\
\noindent $ \textrm{H}_{1}: \beta_{k} \ne 0$.

Typically one conducts one of two different (yet mathematically identical) tests.  Most commonly a $t$-test is done to calculate a $p$-value as follows:
\begin{equation} \label{basicttest}
p\textrm{-value}_{k} = 2\times F_{t}\left(\frac{|
\reallywidehat{\beta}_{k}|}{\textrm{SE}({\reallywidehat{\beta}}_{k})}, N-K-1
\right), \textrm{for $k$ in 0,...,$K$,}
\end{equation}
\noindent where we use $F_{t}(\quad  \cdot \quad ; df)$ to denote the cdf of the $t$-distribution with $df$ degrees of freedom, and where: $\textrm{SE}({\reallywidehat{\beta}}_{k}) = \hat{\sigma}\sqrt{[(X^{T}X)^{-1}]_{kk}}$.  Alternatively, we can conduct an $F$-test and, for $k$ in 1,...,$K$, we will obtain the very same $p$-value with:
\begin{equation} \label{basictestF}
p\textrm{-value}_{k} = p_{F}\left((N-K-1)\frac{\reallywidehat{sr}^{2}_{k}}{1-\hat{R}^{2}_ {Y X}}, 1, N-K-1 \right), 
\end{equation}
\noindent where $p_{f}(\cdot \quad ; df_{1}, df_{2})$ is the cdf of the $F$-distribution with $df_{1}$ and $df_{2}$ degrees of freedom, and where: $ \reallywidehat{sr}^{2}_{k} = \hat{R}^{2}_ {Y X} - \hat{R}^{2}_{Y X_{-k}}$.    Regardless of whether the $t$-test or the $F$-test is employed, if $p\textrm{-value}_{k}<\alpha$, we reject the null hypothesis of $\textrm{H}_{0}:  \beta_{k} = 0 $ against the alternative  $\textrm{H}_{0}:  \beta_{k} \ne 0 $.

\subsubsection*{A valid equivalence test for the standardized difference between two independent means}

 A valid equivalence test for the standardized difference between two independent means, $\theta$, can be defined by the following null and alternative hypotheses (see \citet{serlin1993rational}, \citet{weber2012testing}): 

$\textrm{H}_{0}: \theta \le \Delta_{lower} \quad \textrm{or:}  \quad \theta \ge \Delta_{upper} $, \quad \textrm{vs.}\\
 \indent $\textrm{H}_{1}: \theta > \Delta_{lower} \quad \textrm{and:} \quad \theta < \Delta_{upper}, $

\noindent where $\theta= \mu_{d}/\sigma$ and the equivalence margin is $(\Delta_{lower}, \Delta_{upper})$.  A $p$-value for this test can then be calculated as $p$-value=$\textrm{max}(p_{d}^{lower}, p_{d}^{upper})$, where:
\begin{equation}
    p_{d}^{lower} = 
1-F_{t}\left(\frac{\hat{\mu}_{d}}{\hat{\sigma}_{p}}\sqrt{\frac{N_{1}N_{2}}{N_{1}+N_{2}}}, N_{1}+N_{2}-2, \Delta_{lower}\sqrt{\frac{N_{1}N_{2}}{N_{1}+N_{2}}}\right), \quad \textrm{and}
\label{eq:valid}
\end{equation}
\begin{equation*}
p_{d}^{upper} = 1-F_{t}\left(-\frac{\hat{\mu}_{d}}{\hat{\sigma}_{p}}\sqrt{\frac{N_{1}N_{2}}{N_{1}+N_{2}}}, N_{1}+N_{2}-2, -\Delta_{upper}\sqrt{\frac{N_{1}N_{2}}{N_{1}+N_{2}}}\right),
\end{equation*}
\noindent where $N_{1}$ is the number of observations in the first sample, $N_{2}$ is the number of observations in the second sample,  where $\hat{\mu}_{d}$ is the difference between the two sample means, and $\hat{\sigma}_{p}$, the pooled standard deviation estimate, is calculated from the two samples as:
\begin{equation}
    \hat{\sigma}_{p}=\sqrt{\frac{(N_{1}-1)\hat{\sigma}_{1}^{2} + (N_{2}-1)\hat{\sigma}_{2}^{2} }{N_{1}+N_{2}-2}},
    \label{eq:sp}
\end{equation}
where $\hat{\sigma}_{1}$ is the estimated standard deviation of the first sample, and $\hat{\sigma}_{2}$ is the estimated standard deviation of the second sample.

\subsubsection*{An equivalence test for correlations based on Fisher's Z transformation}

  A $p$-value from the equivalence test for correlations based on Fisher's Z transformation is calculated as $p_{Z}=max(\mathbbmss{p}^{lower}_{Z}, \mathbbmss{p}^{upper}_{Z})$, where:
 \begin{equation}
\mathbbmss{p}^{lower}_{Z} = 1-F_{Z}\left(
\frac{\sqrt{N-3}}{2}\textrm{ln}\left(\left(\frac{1+\reallywidehat{sr}_{1}}{1-\reallywidehat{sr}_{1}}\right) - \left(\frac{1+\Delta_{lower}}{1-\Delta_{lower}}\right)\right)
 \right), \quad \\ \quad
\label{eq:corr2}
\end{equation}
\noindent and:
\begin{equation*}
\mathbbmss{p}^{upper}_{Z} =1-F_{Z}\left(
\frac{\sqrt{N-3}}{2}\textrm{ln}\left(\left(\frac{1+\reallywidehat{sr}_{1}}{1-\reallywidehat{sr}_{1}}\right) + \left(\frac{1+\Delta_{upper}}{1-\Delta_{upper}}\right)\right)
 \right),
\end{equation*}
\noindent where $F_{Z}()$ denotes the cdf of the standard normal distribution; see \citet{goertzen2010detecting} for details.

\pagebreak

\begin{table}[h!!!]
\centering
%\begin{footnotesize}
\begin{tabular}{r|rrrr}
  \hline
$i$ & $X_{1}$ & $X_{2}$ & $X_{3}$ & $Y$ \\ 
 & Received counselling  & Age  & Household income & Anxiety score\\ 
 & (yes=1; no=0)  & (years) & (\$) & points \\ 
 
  \hline
1 & 0 & 13 & 75593 & 12.1 \\ 
  2 & 0 & 15 & 57954 & 15.5 \\ 
  3 & 0 & 13 & 61336 & 13.3 \\ 
  4 & 1 & 14 & 47628 & 14.3 \\ 
  5 & 0 & 14 & 46564 & 12.3 \\ 
  6 & 1 & 12 & 74071 & 13.8 \\ 
  7 & 1 & 17 & 76964 & 14.5 \\ 
  8 & 1 & 15 & 69060 & 11.6 \\ 
  9 & 0 & 13 & 86445 & 12.8 \\ 
  10 & 0 & 17 & 109002 & 17.5 \\ 
  11 & 1 & 16 & 58179 & 14.7 \\ 
  12 & 1 & 14 & 21817 & 16.8 \\ 
  13 & 1 & 17 & 88115 & 12.5 \\ 
  14 & 0 & 17 & 53816 & 15.8 \\ 
  15 & 1 & 16 & 54240 & 17.3 \\ 
  16 & 0 & 16 & 88511 & 15.8 \\ 
  17 & 1 & 16 & 62305 & 16.2 \\ 
  18 & 0 & 15 & 43586 & 13.6 \\ 
  19 & 0 & 14 & 71626 & 12.3 \\ 
  20 & 0 & 14 & 65222 & 12.0 \\ 
  21 & 0 & 14 & 68115 & 14.6 \\ 
  22 & 1 & 15 & 75706 & 13.2 \\ 
  23 & 0 & 13 & 60587 & 12.8 \\ 
  24 & 0 & 19 & 80888 & 16.1 \\ 
  25 & 0 & 17 & 63590 & 20.1 \\ 
  26 & 0 & 13 & 74636 & 12.3 \\ 
  27 & 1 & 14 & 89937 & 15.2 \\ 
  28 & 1 & 14 & 76704 & 15.0 \\ 
  29 & 0 & 16 & 61481 & 13.2 \\ 
  30 & 1 & 15 & 90976 & 15.0 \\ 
  31 & 0 & 15 & 87870 & 17.9 \\ 
  32 & 1 & 15 & 78968 & 16.3 \\ 
  33 & 0 & 15 & 72775 & 14.9 \\ 
  34 & 1 & 17 & 55442 & 15.6 \\ 
  35 & 1 & 15 & 95213 & 10.4 \\ 
  36 & 0 & 18 & 55995 & 19.0 \\ 
  37 & 0 & 12 & 111747 & 9.5 \\ 
  38 & 0 & 16 & 98652 & 16.7 \\ 
  39 & 0 & 15 & 63286 & 19.3 \\ 
  40 & 1 & 15 & 47472 & 12.3 \\ 
   \hline
\end{tabular}
\caption{{The hypothetical anxiety study dataset. In this hypothetical study, $Y$ might be a student's score on an anxiety assessment questionnaire; $X_{1}$ might be a binary variable indicating whether or not the student received counselling services (0 = ``did not receive counselling; 1 = ``did receive counselling''); $X_{2}$ might be a continuous  predictor corresponding to the student's age in years; and $X_{3}$ might be a continuous { predictor }corresponding to the student's household income in dollars.}}
%\end{footnotesize}
\label{tab:anx}
\end{table}

\pagebreak

\subsection*{{Simulation Study 1}}

We simulated data in order to compare the operating characteristics of two equivalence tests for the difference between two independent means:

\begin{enumerate}
    \item the invalid test (i.e., the test proposed by \citet{lakens2017equivalence}), with null hypothesis  $\textrm{H}_{0}: |\mu_{d}| \ge \Delta\times\hat{\sigma}$; and

\item the valid test (see equation (\ref{eq:valid})), with null hypothesis $\textrm{H}_{0}: |\theta| \ge \Delta$, where $\theta=\mu_{d}/\sigma$.
\end{enumerate}

\noindent We considered 6 different values for the total sample size, $N$, ranging from 54 to 3500 (values representative of sample sizes in large and very large psychological studies \citep{kuhberger2014publication, fraley2014n, marszalek2011sample}), and 4 different values for the upper bound of a symmetric equivalence margin, $\Delta$, ranging from 0.2 to 1.0.  We simulated data from a Normal distribution such that the true Cohen's $d$ was equal to 0, or equal to $\Delta$, or equal to 0.15.  

For each of the different configurations within the simulation study, we simulated 2,000,000 unique datasets and calculated a $p$-value with each of the two equivalence tests.  We then calculated the proportion of these $p$-values less than $\alpha=0.05$.  We specifically chose to conduct 2,000,000 simulation runs so as to keep computing time within a reasonable limit while also reducing the amount of Monte Carlo standard error to a very negligible amount (for looking at type 1 error with $\alpha=0.05$, Monte Carlo SE will be approximately $0.00015 \approx \sqrt{0.05(1-0.05)/2,000,000}$; see \cite{morris2019using}).  

 The simulation study was done using the R statistical software with default simulation routines \citep{rsoftware}.  Results are displayed in Table \ref{tab:sim_lakens} and suggest that, in practice, using the invalid test can lead to a higher than advertised type 1 error when sample sizes are large and a minor loss of efficiency when sample sizes are small.

% latex table generated in R 4.1.2 by xtable 1.8-4 package
% Thu Mar  3 15:09:04 2022
\begin{table}[ht]
\centering
\begin{tabular}{|rr|rr|rr|rr|}
  \hline
  $N$ & $\Delta$ & \multicolumn{2}{c|}{$\textrm{Pr}(p\textrm{-val}<0.05 | d = \Delta)$ }  & \multicolumn{2}{c|}{$\textrm{Pr}(p\textrm{-val}<0.05 | d = 0)$} & \multicolumn{2}{c|}{$\textrm{Pr}(p\textrm{-val}<0.05 | d = 0.15)$} \\ 
   &  & invalid test & valid test & invalid test & valid test& invalid test & valid test \\ 
\hline
54 & 0.20 & 0.000 & 0.000 & 0.000 & 0.000 & 0.000 & 0.000 \\ 
  54 & 0.50 & 0.024 & 0.029 & 0.128 & 0.152 & 0.110 & 0.131 \\ 
  54 & 0.75 & 0.047 & 0.050 & 0.716 & 0.727 & 0.647 & 0.658 \\ 
  54 & 1.00 & 0.050 & 0.050 & 0.949 & 0.950 & 0.915 & 0.916 \\ 
  80 & 0.20 & 0.000 & 0.000 & 0.000 & 0.000 & 0.000 & 0.000 \\ 
  80 & 0.50 & 0.045 & 0.048 & 0.430 & 0.443 & 0.352 & 0.364 \\ 
  80 & 0.75 & 0.049 & 0.050 & 0.905 & 0.907 & 0.833 & 0.835 \\ 
  80 & 1.00 & 0.051 & 0.050 & 0.994 & 0.994 & 0.981 & 0.980 \\ 
  180 & 0.20 & 0.000 & 0.000 & 0.000 & 0.000 & 0.000 & 0.000 \\ 
  180 & 0.50 & 0.050 & 0.050 & 0.909 & 0.910 & 0.750 & 0.752 \\ 
  180 & 0.75 & 0.051 & 0.050 & 0.999 & 0.999 & 0.990 & 0.990 \\ 
  180 & 1.00 & 0.054 & 0.050 & 1.000 & 1.000 & 1.000 & 1.000 \\ 
  540 & 0.20 & 0.048 & 0.049 & 0.501 & 0.502 & 0.135 & 0.136 \\ 
  540 & 0.50 & 0.051 & 0.050 & 1.000 & 1.000 & 0.992 & 0.992 \\ 
  540 & 0.75 & 0.053 & 0.050 & 1.000 & 1.000 & 1.000 & 1.000 \\ 
  540 & 1.00 & 0.057 & 0.050 & 1.000 & 1.000 & 1.000 & 1.000 \\ 
  1000 & 0.20 & 0.050 & 0.050 & 0.870 & 0.870 & 0.196 & 0.197 \\ 
  1000 & 0.50 & 0.052 & 0.050 & 1.000 & 1.000 & 1.000 & 1.000 \\ 
  1000 & 0.75 & 0.054 & 0.050 & 1.000 & 1.000 & 1.000 & 1.000 \\ 
  1000 & 1.00 & 0.058 & 0.050 & 1.000 & 1.000 & 1.000 & 1.000 \\ 
  3500 & 0.20 & 0.050 & 0.050 & 1.000 & 1.000 & 0.434 & 0.434 \\ 
  3500 & 0.50 & 0.052 & 0.050 & 1.000 & 1.000 & 1.000 & 1.000 \\ 
  3500 & 0.75 & 0.055 & 0.050 & 1.000 & 1.000 & 1.000 & 1.000 \\ 
  3500 & 1.00 & 0.059 & 0.050 & 1.000 & 1.000 & 1.000 & 1.000 \\ 
   \hline
\end{tabular}
\caption{Results from Simulation Study 1.   Note that the maximum type 1 error rate should not exceed $\alpha=0.05$.  As such, when $\Delta = d$, the probability of a $p$-value less than 0.05 should not exceed 0.05.  When $\Delta>d$, the probability of a $p$-value less than 0.05 corresponds to the test's statistical power.  }
\end{table}
\label{tab:sim_lakens}

\pagebreak

\subsection*{{{Simulation Study 2}}}
We simulated data in order to compare the operating characteristics of two equivalence tests for the difference between two independent means:

\begin{enumerate}

\item the proposed equivalence test for semipartial correlation coefficients (see equation (\ref{eq:spcpval})) (``sr test''); and 
    \item the equivalence test for correlations based on Fisher's Z transformation (see equation (\ref{eq:corr2})) (``Z test'').
\end{enumerate}

\noindent Both tests are valid for testing the lack of an association between $Y$ and $X$ when $K=1$ (i.e., for simple linear regression).  We considered 6 different values for the total sample size, $N$, ranging from 54 to 3500 (values representative of sample sizes in large and very large psychological studies \citep{kuhberger2014publication, fraley2014n, marszalek2011sample}), and 3 different values for the upper bound of a symmetric equivalence margin, $\Delta$, ranging from 0.1 to 0.20.  We simulated data from a bivariate Normal distribution such that the true value of $sr_{1}$ was equal to 0, or equal to $\Delta$, or equal to 0.05.  

For each of the different configurations within the simulation study, we simulated 2,000,000 unique datasets and calculated a $p$-value with each of the two equivalence tests.  We then calculated the proportion of these $p$-values less than $\alpha=0.05$.     We specifically chose to conduct 2,000,000 simulation runs so as to keep computing time within a reasonable limit while also reducing the amount of Monte Carlo standard error to a very negligible amount (for looking at type 1 error with $\alpha=0.05$, Monte Carlo SE will be approximately $0.00015 \approx \sqrt{0.05(1-0.05)/2,000,000}$; see \cite{morris2019using}).  

 The simulation study was done using the R statistical software with default simulation routines \citep{rsoftware}.  Results are displayed in Table \ref{tab:sim_2} and suggest that, in practice, both equivalence tests obtain very similar values for the type 1 error and statistical power.

% latex table generated in R 4.1.2 by xtable 1.8-4 package
% Thu Mar  3 15:09:04 2022
\begin{table}[ht]
\centering
\begin{tabular}{|rr|rr|rr|rr|}
  \hline
  $N$ & $\Delta$ & \multicolumn{2}{c|}{$\textrm{Pr}(p<0.05 | sr_{1} = \Delta)$ }  & \multicolumn{2}{c|}{$\textrm{Pr}(p<0.05 | sr_{1} = 0)$} & \multicolumn{2}{c|}{$\textrm{Pr}(p<0.05 | sr_{1} = 0.05)$}  \\ 
   &  & sr test & Z test & sr test & Z test & sr test & Z test \\ 
\hline
54 & 0.10 & 0.000 & 0.000 & 0.000 & 0.000 & 0.000 & 0.000 \\ 
  54 & 0.15 & 0.000 & 0.000 & 0.000 & 0.000 & 0.000 & 0.000 \\ 
  54 & 0.20 & 0.000 & 0.000 & 0.000 & 0.000 & 0.000 & 0.000 \\ 
  80 & 0.10 & 0.000 & 0.000 & 0.000 & 0.000 & 0.000 & 0.000 \\ 
  80 & 0.15 & 0.000 & 0.000 & 0.000 & 0.000 & 0.000 & 0.000 \\ 
  80 & 0.20 & 0.016 & 0.022 & 0.081 & 0.107 & 0.076 & 0.101 \\ 
  180 & 0.10 & 0.000 & 0.000 & 0.000 & 0.000 & 0.000 & 0.000 \\ 
  180 & 0.15 & 0.038 & 0.041 & 0.273 & 0.286 & 0.237 & 0.249 \\ 
  180 & 0.20 & 0.046 & 0.049 & 0.694 & 0.707 & 0.629 & 0.642 \\ 
  540 & 0.10 & 0.048 & 0.048 & 0.499 & 0.503 & 0.345 & 0.348 \\ 
  540 & 0.15 & 0.048 & 0.050 & 0.935 & 0.937 & 0.817 & 0.820 \\ 
  540 & 0.20 & 0.047 & 0.049 & 0.998 & 0.998 & 0.982 & 0.983 \\ 
  1000 & 0.10 & 0.049 & 0.050 & 0.870 & 0.872 & 0.597 & 0.599 \\ 
  1000 & 0.15 & 0.048 & 0.050 & 0.998 & 0.998 & 0.968 & 0.969 \\ 
  1000 & 0.20 & 0.048 & 0.050 & 1.000 & 1.000 & 1.000 & 1.000 \\ 
  3500 & 0.10 & 0.050 & 0.050 & 1.000 & 1.000 & 0.972 & 0.972 \\ 
  3500 & 0.15 & 0.049 & 0.050 & 1.000 & 1.000 & 1.000 & 1.000 \\ 
  3500 & 0.20 & 0.049 & 0.050 & 1.000 & 1.000 & 1.000 & 1.000 \\ 
   \hline
\end{tabular}
\caption{Results from Simulation Study 2.   Note that the maximum type 1 error rate should not exceed $\alpha=0.05$. When $\Delta>sr_{1}$, the probability of a $p$-value less than 0.05 corresponds to the test's statistical power. The two tests under consideration are the proposed equivalence test for semipartial correlation coefficients (``sr test'') and  the equivalence test for correlations based on Fisher's Z transformation (``Z test''). }
\end{table}
\label{tab:sim_2}

\subsection*{{Simulation Study 3}}
We conducted a simple simulation study in order to better understand the operating characteristics of the proposed equivalence test for the semipartial correlation and to confirm that the proposed formula for approximating statistical power (equation (\ref{power2})) is accurate.  The equivalence test in the simulation study targeted ${sr}_{1}$ and considered a symmetric equivalence margin, $(-\Delta, \Delta)$, such that the hypothesis test in question can be stated as: $\textrm{H}_{0}:  |{sr}_{1}| \ge \Delta$, vs. $\textrm{H}_{1}: |{sr}_{1}| < \Delta$.  

 We considered 4 different values for the total sample size, $N$, ranging from 54 to 3500 (values representative of sample sizes in large and very large psychological studies \citep{kuhberger2014publication, fraley2014n, marszalek2011sample}), and 3 different values for the upper bound of the symmetric equivalence margin, $\Delta$, ranging from 0.10 to 0.20.  We considered two values for $K$, the number of predictors: $K=2$ or $K=4$; and simulated the predictors from a multivariate Normal distribution with a correlation matrix in which all off-diagonal elements were equal to either $\rho_{X}=0.1$ or to $\rho_{X}=0.2$.  Finally, the outcome data, $Y$, was simulated such that the true value of $sr_{1}$ was equal to 0, or equal to $\Delta$, or equal to 0.05. 

For each of the different configurations within the simulation study, we simulated 500,000 unique datasets and calculated a $p$-value with the proposed equivalence test.  We then calculated the proportion of these $p$-values less than $\alpha=0.05$.   We also used the proposed formula for approximating statistical power for each scenario to calculate the approximate power.  We specifically chose to conduct 500,000 simulation runs so as to keep computing time within a reasonable limit while also reducing the amount of Monte Carlo standard error to a very negligible amount (for looking at type 1 error with $\alpha=0.05$, Monte Carlo SE will be approximately $0.0003 \approx \sqrt{0.05(1-0.05)/500,000}$; see \cite{morris2019using}).  

 The simulation study was done using the R statistical software with default simulation routines \citep{rsoftware}.  Results are displayed in Table \ref{tab:sim_3} and suggest that, in practice, the proposed test (``sr test'') has correct type 1 error and that the proposed formula for estimating statistical power (``approx pwr.'') is reasonably accurate.

% latex table generated in R 4.1.2 by xtable 1.8-4 package
% Thu Mar  3 15:09:04 2022
\begin{table}[h!!]
\centering
\begin{footnotesize}
\begin{tabular}{|rrrr|rr|rr|rr|}
  \hline
  $N$ & $\Delta$ & $K$ & $\rho_{X}$& \multicolumn{2}{c|}{$\textrm{Pr}(p<0.05 | sr_{1} = \Delta)$ }  & \multicolumn{2}{c|}{$\textrm{Pr}(p<0.05 | sr_{1} = 0)$} & \multicolumn{2}{c|}{$\textrm{Pr}(p<0.05 | sr_{1} = 0.05)$}  \\ 
  && &  & sr test & approx pwr. & sr test & approx pwr. & sr test & approx pwr. \\ 
\hline
54 & 0.10 & 2.0 & 0.10 & 0.000 & 0.000 & 0.000 & 0.000 & 0.000 & 0.000 \\ 
  54 & 0.15 & 2.0 & 0.10 & 0.000 & 0.000 & 0.000 & 0.000 & 0.000 & 0.000 \\ 
  54 & 0.20 & 2.0 & 0.10 & 0.004 & 0.000 & 0.000 & 0.000 & 0.000 & 0.000 \\ 
  180 & 0.10 & 2.0 & 0.10 & 0.000 & 0.000 & 0.000 & 0.000 & 0.000 & 0.000 \\ 
  180 & 0.15 & 2.0 & 0.10 & 0.044 & 0.046 & 0.303 & 0.301 & 0.225 & 0.229 \\ 
  180 & 0.20 & 2.0 & 0.10 & 0.049 & 0.050 & 0.718 & 0.713 & 0.601 & 0.598 \\ 
  540 & 0.10 & 2.0 & 0.10 & 0.049 & 0.049 & 0.528 & 0.527 & 0.284 & 0.285 \\ 
  540 & 0.15 & 2.0 & 0.10 & 0.049 & 0.050 & 0.945 & 0.942 & 0.756 & 0.754 \\ 
  540 & 0.20 & 2.0 & 0.10 & 0.049 & 0.050 & 0.998 & 0.998 & 0.970 & 0.968 \\ 
  3500 & 0.10 & 2.0 & 0.10 & 0.050 & 0.050 & 1.000 & 1.000 & 0.910 & 0.909 \\ 
  3500 & 0.15 & 2.0 & 0.10 & 0.050 & 0.050 & 1.000 & 1.000 & 1.000 & 1.000 \\ 
  3500 & 0.20 & 2.0 & 0.10 & 0.050 & 0.050 & 1.000 & 1.000 & 1.000 & 1.000 \\ 
  54 & 0.10 & 4.0 & 0.10 & 0.000 & 0.000 & 0.000 & 0.000 & 0.000 & 0.000 \\ 
  54 & 0.15 & 4.0 & 0.10 & 0.000 & 0.000 & 0.000 & 0.000 & 0.000 & 0.000 \\ 
  54 & 0.20 & 4.0 & 0.10 & 0.003 & 0.000 & 0.000 & 0.000 & 0.000 & 0.000 \\ 
  180 & 0.10 & 4.0 & 0.10 & 0.000 & 0.000 & 0.000 & 0.000 & 0.000 & 0.000 \\ 
  180 & 0.15 & 4.0 & 0.10 & 0.044 & 0.045 & 0.299 & 0.299 & 0.225 & 0.227 \\ 
  180 & 0.20 & 4.0 & 0.10 & 0.050 & 0.050 & 0.717 & 0.711 & 0.600 & 0.596 \\ 
  540 & 0.10 & 4.0 & 0.10 & 0.048 & 0.049 & 0.525 & 0.524 & 0.282 & 0.284 \\ 
  540 & 0.15 & 4.0 & 0.10 & 0.049 & 0.050 & 0.943 & 0.941 & 0.756 & 0.753 \\ 
  540 & 0.20 & 4.0 & 0.10 & 0.049 & 0.050 & 0.998 & 0.998 & 0.971 & 0.968 \\ 
  3500 & 0.10 & 4.0 & 0.10 & 0.049 & 0.050 & 1.000 & 1.000 & 0.909 & 0.909 \\ 
  3500 & 0.15 & 4.0 & 0.10 & 0.049 & 0.050 & 1.000 & 1.000 & 1.000 & 1.000 \\ 
  3500 & 0.20 & 4.0 & 0.10 & 0.049 & 0.050 & 1.000 & 1.000 & 1.000 & 1.000 \\ 
  54 & 0.10 & 2.0 & 0.25 & 0.000 & 0.000 & 0.000 & 0.000 & 0.000 & 0.000 \\ 
  54 & 0.15 & 2.0 & 0.25 & 0.000 & 0.000 & 0.000 & 0.000 & 0.000 & 0.000 \\ 
  54 & 0.20 & 2.0 & 0.25 & 0.007 & 0.010 & 0.000 & 0.000 & 0.000 & 0.000 \\ 
  180 & 0.10 & 2.0 & 0.25 & 0.000 & 0.000 & 0.000 & 0.000 & 0.000 & 0.000 \\ 
  180 & 0.15 & 2.0 & 0.25 & 0.045 & 0.047 & 0.303 & 0.301 & 0.227 & 0.231 \\ 
  180 & 0.20 & 2.0 & 0.25 & 0.049 & 0.050 & 0.718 & 0.713 & 0.602 & 0.600 \\ 
  540 & 0.10 & 2.0 & 0.25 & 0.049 & 0.049 & 0.528 & 0.527 & 0.285 & 0.286 \\ 
  540 & 0.15 & 2.0 & 0.25 & 0.049 & 0.050 & 0.945 & 0.942 & 0.757 & 0.756 \\ 
  540 & 0.20 & 2.0 & 0.25 & 0.049 & 0.050 & 0.998 & 0.998 & 0.971 & 0.969 \\ 
  3500 & 0.10 & 2.0 & 0.25 & 0.050 & 0.050 & 1.000 & 1.000 & 0.911 & 0.910 \\ 
  3500 & 0.15 & 2.0 & 0.25 & 0.050 & 0.050 & 1.000 & 1.000 & 1.000 & 1.000 \\ 
  3500 & 0.20 & 2.0 & 0.25 & 0.050 & 0.050 & 1.000 & 1.000 & 1.000 & 1.000 \\ 
  54 & 0.10 & 4.0 & 0.25 & 0.000 & 0.000 & 0.000 & 0.000 & 0.000 & 0.000 \\ 
  54 & 0.15 & 4.0 & 0.25 & 0.000 & 0.000 & 0.000 & 0.000 & 0.000 & 0.000 \\ 
  54 & 0.20 & 4.0 & 0.25 & 0.002 & 0.000 & 0.000 & 0.000 & 0.000 & 0.000 \\ 
  180 & 0.10 & 4.0 & 0.25 & 0.000 & 0.000 & 0.000 & 0.000 & 0.000 & 0.000 \\ 
  180 & 0.15 & 4.0 & 0.25 & 0.043 & 0.045 & 0.295 & 0.295 & 0.225 & 0.226 \\ 
  180 & 0.20 & 4.0 & 0.25 & 0.049 & 0.050 & 0.714 & 0.708 & 0.600 & 0.595 \\ 
  540 & 0.10 & 4.0 & 0.25 & 0.048 & 0.049 & 0.521 & 0.521 & 0.282 & 0.283 \\ 
  540 & 0.15 & 4.0 & 0.25 & 0.049 & 0.050 & 0.942 & 0.940 & 0.756 & 0.752 \\ 
  540 & 0.20 & 4.0 & 0.25 & 0.049 & 0.050 & 0.998 & 0.998 & 0.971 & 0.968 \\ 
  3500 & 0.10 & 4.0 & 0.25 & 0.049 & 0.050 & 1.000 & 1.000 & 0.909 & 0.908 \\ 
  3500 & 0.15 & 4.0 & 0.25 & 0.049 & 0.050 & 1.000 & 1.000 & 1.000 & 1.000 \\ 
  3500 & 0.20 & 4.0 & 0.25 & 0.049 & 0.050 & 1.000 & 1.000 & 1.000 & 1.000 \\ 
   \hline
\end{tabular}
\end{footnotesize}
\caption{Results from Simulation Study 3.   Note that the maximum type 1 error rate should not exceed $\alpha=0.05$.  When $\Delta>sr_{1}$, the probability of a $p$-value less than 0.05 corresponds to the test's statistical power.  }
\end{table}
\label{tab:sim_3}

\pagebreak

\vspace{10cm}
\subsection*{R code}
\begin{footnotesize}
    
All data and code has been saved in csv format and is available in the OSF repository at https://osf.io/5yr92/, DOI 10.17605/OSF.IO/5YR92.

\begin{verbatim}
############################################
## Useful functions
############################################
equiv_corrZ<-function(var1, var2, delta_upper, delta_lower = NA) {
  
  if(is.na(delta_lower)){delta_lower <-(-delta_upper)}
  
  corxy<-cor(var1,var2)
  n<-length(var1)
  
  ##### Run a two t-test procedure for equivalance with Fisher's z transformation ####
  zei_lower <- log((1-delta_lower)/(1+delta_lower))/2
  zei_upper <- log((1+delta_upper)/(1-delta_upper))/2
  zcorxy<-log((1+corxy)/(1-corxy))/2
  equivt1_fz<-(zcorxy+ zei_lower)/(1/sqrt(n-3))
  pvalue1_fz<-1-pnorm(equivt1_fz)
  equivt2_fz<-(zcorxy- zei_upper)/(1/sqrt(n-3))
  pvalue2_fz<-pnorm(equivt2_fz)
  
  the_results <- c(pvalue_equiv_z=max(c(pvalue1_fz, pvalue2_fz), na.rm=TRUE))
  return(the_results)
}
############################################

############################################
equivBeta <- function(Y = rnorm(100), 
                      Xmatrix = cbind(rnorm(100), rnorm(100)), 
                      DELTA_upper = 0.1,
                      DELTA_lower = -0.1){
  if(is.na(DELTA_lower)[1]){DELTA_lower <-(-DELTA_upper)}                      
  Xmatrix <- cbind(Xmatrix)	
  X <- cbind(1, Xmatrix)
  N <- dim(cbind(X[,-1]))[1]
  K <- dim(cbind(X[,-1]))[2]
  if(length(DELTA_lower)==1){DELTA_lower <- rep(DELTA_lower, K+1)}
  if(length(DELTA_upper)==1){DELTA_upper <- rep(DELTA_upper, K+1)}
  
  lmmod <- summary(lm(Y~X[,-1]))
  beta_hat <- lmmod$coef[,1]
  SE_beta_hat <- lmmod$coef[,2]
  
  mysigma<-summary(lm(Y~X[,-1]))$sigma
  mysigma*sqrt(solve(t(X)%*%X)[k,k])
  
  pval <- p_lower <- p_upper <- rep(0,K)
  for(k in 1:(K+1)){ 
    p_lower[k] <- pt((beta_hat[k] - DELTA_lower[k])/SE_beta_hat[k], N-K-1, 0, 
                     lower.tail=FALSE)
    p_upper[k] <- pt((-beta_hat[k] + DELTA_upper[k])/SE_beta_hat[k], N-K-1, 0, 
                     lower.tail=FALSE)
    pval[k] <- max(c(p_lower[k],p_upper[k]))
  }
  
  names(beta_hat) <- paste("beta", c(1:dim(X)[2])-1, sep="_")
  names(pval) <- paste("pval", c(1:dim(X)[2])-1, sep="_")
  DELTA = cbind(DELTA_lower, DELTA_upper)
  rownames(DELTA) <- paste("DELTA", c(1:dim(X)[2])-1, sep="_")
  return(list(beta = beta_hat, pval = pval, DELTA = DELTA))
}
############################################

############################################
equivBetaPower <- function(DELTA_upper, DELTA_lower, N, K, SEbetak, true_beta=0){
  ncp1 <- (DELTA_upper-true_beta)/SEbetak
  ncp2 <- (DELTA_lower-true_beta)/SEbetak
  Tstatstar <- qt(1-0.05, N-K-1)
  power = pt(+ncp1-Tstatstar, N-K-1, lower.tail=TRUE) - pt(+ncp2+Tstatstar, 
              N-K-1, lower.tail=TRUE)
  return(power)
}
############################################

############################################
equivSR <- function(Y= rnorm(100), 
         Xmatrix= cbind(rnorm(100),rnorm(100)), 
         DELTA_upper= 0.1,
         DELTA_lower= NA){
  
  Xmatrix <-cbind(Xmatrix)		
  X <- cbind(1,Xmatrix)	
  N <- dim(Xmatrix)[1]
  K <- dim(Xmatrix)[2]
  kvec= 1:K
  if(is.na(DELTA_lower[1])){DELTA_lower <-(-DELTA_upper)}
  if(length(DELTA_upper)!=K){DELTA_upper <- rep(DELTA_upper[1], K)}
  if(length(DELTA_lower)!=K){DELTA_lower <- rep(DELTA_lower[1], K)}
  
  lmmod <- summary(lm(Y~X[,-1]))
  R2 <- lmmod$r.squared
  if(K==1){R2Xkmink <- 0; diffR2k<-R2; R2Ymink<-0}
  if(K>1){
    R2Ymink <- apply(cbind(kvec),1,function(k)summary(lm(Y~Xmatrix[,-k]))$r.squared)
    R2Xkmink <- apply(cbind(kvec),1,function(k)summary(lm(Xmatrix[,k]~Xmatrix[,-k]))$r.squared)
    diffR2k <- unlist(lapply(c(kvec), function(k) {R2-summary(lm(Y~Xmatrix[,-k]))$r.squared}))
  }
  lmmod_scale <- summary(lm(scale(Y)~scale(X[,-1])-1))
  SPC <- lmmod_scale$coef[,1]*sqrt(1-R2Xkmink)
  # should be equal in abs:
  c(sqrt(diffR2k),SPC)
  
  # equation (11) of Dudgeon (2016)
  SIGMA2_SPC <- (R2^2 - 2*R2 + R2Ymink + 1 - R2Ymink^2)/(N-K-1)
  SE_SPC <- sqrt(SIGMA2_SPC)
  
  CI90_upper_squared <- CI95_upper <- CI95_lower <- CI90_upper <- 
    CI90_lower <- pval <- pval1 <- pval2 <- rep(0, length(kvec))
  for(k in kvec){
    pval1[k] <- pt((SPC[k]-DELTA_lower[k])/SE_SPC[k], N-K-1, lower.tail=FALSE)
    pval2[k] <- pt((DELTA_upper[k]-SPC[k])/SE_SPC[k], N-K-1, lower.tail=FALSE)
    CI90_upper[k] <- SPC[k] - qt(0.05,df=N-K-1)* SE_SPC[k]
    CI90_lower[k] <- SPC[k] + qt(0.05,df=N-K-1)* SE_SPC[k]
    
    CI95_upper[k] <- SPC[k] - qt(0.025,df=N-K-1)* SE_SPC[k]
    CI95_lower[k] <- SPC[k] + qt(0.025,df=N-K-1)* SE_SPC[k]
    
    CI90_upper_squared[k] <- (SPC[k] - qt(0.1,df=N-K-1)* SE_SPC[k])^2
    pval[k] <- max(c(pval1[k], pval2[k]))
  }
  
  CI90 <- cbind(CI90_lower, CI90_upper)
  CI95 <- cbind(CI95_lower, CI95_upper)
  LEAD <- apply(abs(CI90),1,max)
  
  names(SPC) <- paste("sr", 1:K, sep="_")
  names(SE_SPC) <- paste("SE_sr", 1:K, sep="_")
  names(pval) <- paste("pval", 1:K, sep="_")
  rownames(CI90) <- paste("CI90", 1:K, sep="_")
  rownames(CI95) <- paste("CI95", 1:K, sep="_")
  names(LEAD) <- paste("LEAD", 1:K, sep="_")
  return(list(sr = unname(SPC), SE_sr= SE_SPC , pval= pval,
              CI90=CI90, CI95=CI95, LEAD=LEAD))
}
############################################

############################################
equivSRPower <- function(DELTA_upper, DELTA_lower, N, K, SESRk, true_SR=0){
  ncp1 <- (DELTA_upper-true_SR)/SESRk
  ncp2 <- (DELTA_lower-true_SR)/SESRk
  Tstatstar <- qt(1-0.05, N-K-1)
  power = pt(+ncp1-Tstatstar, N-K-1, lower.tail=TRUE) - pt(+ncp2+Tstatstar, 
            N-K-1, lower.tail=TRUE)
  return(power)
}
############################################

############################################
BFstandardBeta <- function(Y= yvec, Xmatrix= Xmat, BFthres=3, random=FALSE){
  
  K<-dim(Xmatrix)[2]
  mydata<-data.frame(Y, Xmatrix)
  colnames(mydata) <- c(c("yvector"),paste("X",1:K,sep=""))
  BFmod <- regressionBF(yvector~. , data= mydata)
  
  BF<-result<-rep(0,K)
  for(k in 1:K){
    whichk<-paste("X",k,sep="")
    BF_without_k <-BFmod[!grepl(whichk,names(BFmod)$numerator)][
      which.max(nchar(names(BFmod)$numerator[!grepl(whichk,names(BFmod)$numerator)]))]
    BF_full <- BFmod[which.max(nchar(names(BFmod)$numerator))]
    BF[k] <- exp(as.numeric(slot(BF_without_k, 
                "bayesFactor")[1]))/exp(as.numeric(slot(BF_full,"bayesFactor")[1]))
    if(BF[k]<= 1/BFthres){result[k]<-"positive" }
    if(BF[k]>BFthres){result[k]<-"negative"}	
    if(BF[k]> 1/BFthres & BF[k]<BFthres){result[k]<-"inconclusive"}
  }
  return(list(BF=c(BF), BFthres=c(BFthres),conclusion= result))
}
############################################

############################################
## Hypothetical anxiety study example:
############################################
equivBetaPower(DELTA_upper=2, DELTA_lower=-2, N=40, K=3, SEbetak=0.63)
#0.8540956

# Note: hypothetical data was created with the following code:
#set.seed(123)
#anx <- cbind(1,sample(c(0,1),40,TRUE),round(rnorm(40,15,1.8)), round(rnorm(40,68000,20000)))
#score <- round(anx%*%c(8,0.6,0.5,-0.00001) + rnorm(40,0,2.3),1)

anxiety_data <- read.csv("anxiety.csv")[,-1]
score <- anxiety_data[,1]
anx <- as.matrix(anxiety_data[,-1])

# Study results:
coefficients(summary(lm(score~anx-1)))[,1:2]
confint((lm(score~anx-1)), level=0.95)
confint((lm(score~anx-1)), level=0.90)

# Equivalence test for regression coef:
equivBeta(Y = score, Xmatrix = anx[,-1], DELTA_lower = -2, DELTA_upper = 2)

# or "by hand":
beta_hat <- coefficients(summary(lm(score~anx-1)))[2,1]
SE_beta_hat <- coefficients(summary(lm(score~anx-1)))[2,2]
DELTA_lower <- -2
DELTA_upper <- 2
N <- 40
K <- 3
p_lower <- pt((beta_hat - DELTA_lower)/SE_beta_hat, N-K-1, 0, lower.tail=FALSE)
p_upper <- pt((-beta_hat + DELTA_upper)/SE_beta_hat, N-K-1, 0, lower.tail=FALSE)
pval <- max(c(p_lower,p_upper))
pval
# 0.017733

# LEAD (or "equivalence confidence interval"):
# 90% CI:
CI90 <- c(beta_hat-qt(1-0.05,N-K-1)*SE_beta_hat, beta_hat+qt(1-0.05,N-K-1)*SE_beta_hat)
LEAD <- max(abs(CI90))
LEAD
#  1.674596
equivBeta(Y = score, Xmatrix = anx[,-1], DELTA_lower = -LEAD, DELTA_upper = LEAD)$pval[2]
#  0.05 

############################################
## Salaries example
############################################
Salaries <- read.csv("Salaries.csv")[,-1]
# or obtain data from: 
# library(carData)

### simple linear regression:
y <- Salaries$salary
X <- model.matrix(lm(salary ~ sex, data=Salaries))

summary(lm(salary ~ sex, data=Salaries))$coef[1:2,1:2]
# Estimate Std. Error
# (Intercept) 101002.41   4809.386
# sexMale      14088.01   5064.579
summary(lm(salary ~ sex, data=Salaries))$sigma
# 30034.61
salaries_equiv <- equivSR(Y = y, Xmatrix = X[,-1], DELTA_upper = 0.5, DELTA_lower = -0.5)
salaries_equiv["sr"]
# 0.1386102
salaries_equiv["SE_sr"]
# 0.04934876
equivBeta(Y = y, Xmatrix = X[,-1], DELTA_upper = 5000,  DELTA_lower = -5000)$pval[2]
# 0.9632451

# the same equivalence test done "by hand":
beta_hat <- coefficients(summary(lm(salary~sex,data=Salaries)))[2,1]
SE_beta_hat <- coefficients(summary(lm(salary~sex,data=Salaries)))[2,2]
DELTA_lower <- -5000
DELTA_upper <- 5000
N <- 397
K <- 1
p_lower <- pt((beta_hat - DELTA_lower)/SE_beta_hat, N-K-1, 0, lower.tail=FALSE)
p_upper <- pt((-beta_hat + DELTA_upper)/SE_beta_hat, N-K-1, 0, lower.tail=FALSE)
pval <- max(c(p_lower,p_upper))
pval

equivSR(Y = y, Xmatrix = X[,-1], DELTA_upper = 0.1, DELTA_lower = -0.1)$pval
# 0.7827741
# the same equivalence test done "by hand":
p_lower <- pt((sr1- (-0.10))/SEsr1, 397-1-1, lower.tail=FALSE)
p_upper <- pt((0.10-sr1)/SEsr1,397-1-1, lower.tail=FALSE)
pval <- max(c(p_lower,p_upper))
pval

library(BayesFactor)
sdata <- data.frame(salary = Salaries$salary, 
            sex = as.numeric(as.factor(Salaries$sex)) - 1)
regressionBF(salary ~ sex, data = sdata)
# 4.525
linearReg.R2stat(N = 397, p = 1, R2 = summary(lm(salary ~ sex, 
        data=Salaries))$r.squared, simple = TRUE)
# 4.525
lmBF(salary ~ sex, data = Salaries)
#  6.177
lmBF(salary ~ sex, data = sdata)
# 4.525

### multiple linear regression:
y <- Salaries$salary
X <- model.matrix(lm(salary ~ sex + yrs.since.phd  + yrs.service + discipline + rank, 
                     data=Salaries))

# NHST p-values
mod1 <- summary(lm(salary ~ sex + yrs.since.phd  + yrs.service + discipline + 
           rank, data=Salaries))
mod1$coef[-1,4]
# 2.158412e-01  2.697855e-02  2.142543e-02  1.878412e-09  1.983251e-03  2.296130e-23
mod1$coef[,1:2]

SPCobj <- equivSR(Y = y, Xmatrix = X[,-1], DELTA_upper = 0.1, DELTA_lower = -0.1)
round(SPCobj$pval,4)
# 0.0761 0.3249 0.3577 0.9997 0.6699 1.0000 
round(SPCobj$LEAD,4)
# 0.1080 0.1446 0.1480 0.2913 0.1780 0.3999 
mod1$sigma
# 22538.65
mod1$r.squared
# 0.4546766

# Bayes Factors
BFs <- (BFstandardBeta(Y= y, Xmatrix=X[,-1])$BF)
BFs
# 3.860594e+00 7.352492e-01 6.036516e-01 1.540123e-07 7.331954e-02 5.592721e-21
round(cbind(mod1$coef[-1,4],SPCobj$pval,BFs,1/BFs),3)
#                            BFs             
#sexMale       0.216 0.076 3.861 2.590000e-01
#yrs.since.phd 0.027 0.325 0.735 1.360000e+00
#yrs.service   0.021 0.358 0.604 1.657000e+00
#disciplineB   0.000 1.000 0.000 6.492987e+06
#rankAsstProf  0.002 0.670 0.073 1.363900e+01
#rankProf      0.000 1.000 0.000 1.138788e+15
############################################
## Mindset Theory example
############################################
mind <- read.csv("Mindset.csv")
mind[,"cog"]<-rowMeans(cbind(scale(mind[,"Cattell.Score"]), scale(mind[,"Letter.Sets.Score"])))

# Testing Premise 1: people with growth mind-sets hold learning goals
coefficients(summary(lm(X1.Learning.Goal~Mindset.Score, data=mind)))[2,]
Z <- equiv_corrZ(mind[,"X1.Learning.Goal"], mind[,"Mindset.Score"], 0.2, -Inf)
S <- equivSR(mind[,"X1.Learning.Goal"],
        mind[,"Mindset.Score"], DELTA_upper=0.2, DELTA_lower=-Inf)

res1 <- c((S$sr)[1], Z, S$pval)
res1
# 0.09774130     0.01450918     0.01582211 

# Testing Premise 2: people with fixed mind-sets hold performance goals
coefficients(summary(lm(X2.Performance.Goal~Mindset.Score, data=mind)))[2,]
Z <- equiv_corrZ(mind[,"X2.Performance.Goal"], mind[,"Mindset.Score"], Inf, -0.2)
S <- equivSR(mind[,"X2.Performance.Goal"],  
      mind[,"Mindset.Score"], DELTA_upper=Inf, DELTA_lower=-0.20)
res2 <- c((S$sr)[1], Z, S$pval)
res2
#  -0.10894803     0.02576879     0.02749891

# Testing Premise 3: people with fixed mind- sets hold performance-avoidance goals
coefficients(summary(lm(X3.Performance.Avoidance.Goal~Mindset.Score, data=mind)))[2,]
Z <- equiv_corrZ(mind[,"X3.Performance.Avoidance.Goal"], mind[,"Mindset.Score"], Inf, -0.2)
S <- equivSR(mind[,"X3.Performance.Avoidance.Goal"], 
      mind[,"Mindset.Score"], DELTA_upper=Inf, DELTA_lower=-0.20)
res3 <- c((S$sr)[1], Z, S$pval)
res3
#  -0.0391385402   0.0003229064   0.0004179885

# Testing Premise 4: people with fixed mind-sets believe 
# that talent alone— without effort—creates success
coefficients(summary(lm(X4.Belief.in.Talent~Mindset.Score, data=mind)))[2,]
Z <- equiv_corrZ(mind[,"X4.Belief.in.Talent"], mind[,"Mindset.Score"], Inf, -0.2)
S <- equivSR(mind[,"X4.Belief.in.Talent"], 
       mind[,"Mindset.Score"], DELTA_upper=Inf, DELTA_lower=-0.20)
res4 <- c((S$sr)[1], Z, S$pval)
res4
#  -0.061215284    0.001588981    0.001906768

# Testing Premise 5: people with growth mind-sets persist to overcome challenges
coefficients(summary(lm(X5.Response.To.Challenge~Mindset.Score, data=mind)))[2,]
Z <- equiv_corrZ(mind[,"X5.Response.To.Challenge"], mind[,"Mindset.Score"], 0.2, -Inf)
S <- equivSR(mind[,"X5.Response.To.Challenge"],
      mind[,"Mindset.Score"], DELTA_upper=0.20, DELTA_lower=-Inf)
res5 <- c((S$sr)[1], Z, S$pval)
res5
#  0.055873780    0.001100171    0.001343043 

# Testing Premise 6: people with growth mind-sets are more resilient following failure
coefficients(summary(lm(X6.Raven.Test.Score~Mindset.Score, data=mind)))[2,]
Z <- equiv_corrZ(mind[,"X6.Raven.Test.Score"], mind[,"Mindset.Score"], 0.2, -Inf)
S <- equivSR(mind[,"X6.Raven.Test.Score"],
      mind[,"Mindset.Score"], DELTA_upper=0.20, DELTA_lower=-Inf)
res6 <- c((S$sr)[1], Z, S$pval)
res6
#  -1.217740e-01   5.976098e-12   1.525199e-11

# Testing Premise 6a: people with growth mind-sets are 
# more resilient following failure when controlling for cognitive ability
summary(lm(X6.Raven.Test.Score~Mindset.Score+ cog, data=mind))
Z1 <- NA
S <- equivSR(mind[,"X6.Raven.Test.Score"],
       as.matrix(mind[,c("Mindset.Score", "cog")]), DELTA_upper=0.20, DELTA_lower=-Inf)
res6a <- c(S$sr[1], Z1, S$pval[1])
res6a
# -5.492464e-02            NA  1.109186e-09  

mindset_results <- round(rbind(res1, res2, res3, res4, res5, res6, res6a), digits=3)
mindset_results
#              pvalue_equiv_z pval_1
# res1   0.098          0.015  0.016
# res2  -0.109          0.026  0.027
# res3  -0.039          0.000  0.000
# res4  -0.061          0.002  0.002
# res5   0.056          0.001  0.001
# res6  -0.122          0.000  0.000
# res6a -0.055             NA  0.000



\end{verbatim}
\end{footnotesize}

\end{document}